\documentstyle[preprint,aps,epsfig]{revtex}

\begin{document}
\begin{flushright}CEBAF-TH-94-27\end{flushright}\vspace{2.0 cm}
\def\bra#1{{\langle#1\vert}}
\def\ket#1{{\vert#1\rangle}}
\def\coeff#1#2{{\scriptstyle\frac{#1}{ #2}}}
\def\undertext#1{{$\underline{\hbox{#1}}$}}
\def\hcal#1{{\hbox{\cal #1}}}
\def\sst#1{{\scriptscriptstyle #1}}
\def\eexp#1{{\hbox{e}^{#1}}}
\def\rbra#1{{\langle #1 \vert\!\vert}}
\def\rket#1{{\vert\!\vert #1\rangle}}
\def\lsim{{ <\atop\sim}}
\def\gsim{{ >\atop\sim}}
\def\refmark#1{{$^{\hbox{#1}}$}}
\def\nubar{{\bar\nu}}
\def\Gmu{{G_\mu}}
\def\alr{{A_\sst{LR}}}
\def\wpv{{W^\sst{PV}}}
\def\evec{{\vec e}}
\def\notq{{\not\! q}}
\def\notk{{\not\! k}}
\def\notp{{\not\! p}}
\def\notpp{{\not\! p'}}
\def\notder{{\not\! \partial}}
\def\notcder{{\not\!\! D}}
\def\Jem{{J_\mu^{em}}}
\def\Jana{{J_{\mu 5}^{anapole}}}
\def\nue{{\nu_e}}
\def\mn{{m_\sst{N}}}
\def\mns{{m^2_\sst{N}}}
\def\me{{m_e}}
\def\mes{{m^2_e}}
\def\mmu{{m_\mu}}
\def\mmus{{m^2_\mu}}
\def\mf{{m_f}}
\def\mfs{{m_f^2}}
\def\mfp{{m_{f'}}}
\def\mfps{{m_{f'}^2}}
\def\mq{{m_q}}
\def\mqs{{m_q^2}}
\def\ml{{m_\ell}}
\def\mls{{m_\ell^2}}
\def\mt{{m_t}}
\def\mts{{m_t^2}}
\def\mnu{{m_\nu}}
\def\mnus{{m_\nu^2}}
\def\mz{{M_\sst{Z}}}
\def\mzs{{M^2_\sst{Z}}}
\def\mw{{M_\sst{W}}}
\def\mws{{M^2_\sst{W}}}
\def\mh{{M_\sst{H}}}
\def\mhs{{M^2_\sst{H}}}
\def\mzb{\frac{\mzs}{\mws}}
\def\mhz{\frac{\mhs}{\mzs}}
\def\mhw{\frac{\mhs}{\mws}}
\def\mfw{\frac{\mfs}{\mws}}
\def\ubar{{\bar u}}
\def\dbar{{\bar d}}
\def\sbar{{\bar s}}
\def\qbar{{\bar q}}
\def\Abar{{\overline A}}
\def\Nbar{{\overline N}}
\def\ucr{{u^{\dag}}}
\def\dcr{{d^{\dag}}}
\def\QM{{\sst{QM}}}
\def\ctw{{\cos\theta_\sst{W}}}
\def\stw{{\sin\theta_\sst{W}}}
\def\sstw{{\sin^2\theta_\sst{W}}}
\def\cstw{{\cos^2\theta_\sst{W}}}
\def\cftw{{\cos^4\theta_\sst{W}}}
\def\tw{{\theta_\sst{W}}}
\def\sstwh{{\sin^2{\hat\theta}_\sst{W}}}
\def\sstwb{{\sin^2{\bar\theta}_\sst{W}}}
\def\ztil{{{\tilde Z}^{1/2}}}
\def\ztilij{{{\tilde Z}^{1/2}_{ij}}}
\def\zstil{{\tilde Z}}
\def\zren{{Z^{1/2}}}
\def\zrenw{{Z^{1/2}_\sst{WW}}}
\def\zrenz{{Z^{1/2}_\sst{ZZ}}}
\def\zrena{{Z^{1/2}_\sst{AA}}}
\def\zrenaz{{Z^{1/2}_\sst{AZ}}}
\def\zrenza{{Z^{1/2}_\sst{ZA}}}
\def\zrenl{{Z^{1/2}_\sst{L}}}
\def\zrenr{{Z^{1/2}_\sst{R}}}
\def\zrenps{{Z^{1/2}_\psi}}
\def\zrenpsb{{Z^{1/2}_{\bar\psi}}}
\def\znren{{Z^{-1/2}}}
\def\dw{{\delta_\sst{W}}}
\def\dz{{\delta_\sst{Z}}}
\def\dzb{{{\overline\delta}_\sst{Z}}}
\def\da{{\delta_\sst{A}}}
\def\dza{{\delta_\sst{ZA}}}
\def\dzap{{\delta^{pole}_\sst{ZA}}}
\def\dzab{{{\overline\delta}_\sst{ZA}}}
\def\daz{{\delta_\sst{AZ}}}
\def\dazp{{\delta^{pole}_\sst{AZ}}}
\def\dazb{{{\overline\delta}_\sst{AZ}}}
\def\dmw{{\delta M^2_\sst{W}}}
\def\dmz{{\delta M^2_\sst{Z}}}
\def\dmwb{{{\overline\dmw}}}
\def\dmzb{{{\overline\dmz}}}
\def\dy{{\delta_\sst{Y}}}
\def\dyb{{{\overline\delta}_\sst{Y}}}
\def\dps{{\delta_\psi}}
\def\dpsf{{\delta_\psi^5}}
\def\dl{{\delta_\sst{L}}}
\def\dr{{\delta_\sst{R}}}
\def\tmunu{{T_{\mu\nu}}}
\def\lmunu{{L_{\mu\nu}}}
\def\gp{{(\xi-1)}}
\def\cc{\frac{\alpha}{ (4\pi)}}
\def\auv{{\alpha_\sst{UV}}}
\def\air{{\alpha_\sst{IR}}}
\def\qw{{Q_\sst{W}^2}}
\def\Gf{{G_\sst{F}}}
\def\gv{{g_\sst{V}}}
\def\ga{{g_\sst{A}}}
\def\gvq{{g_\sst{V}^q}}
\def\gaq{{g_\sst{A}^q}}
\def\gvf{{g_\sst{V}^{f}}}
\def\gaf{{g_\sst{A}^{f}}}
\def\gvfp{{g_\sst{V}^{f'}}}
\def\gafp{{g_\sst{A}^{f'}}}
\def\gvfs{{{\gvf}^2}}
\def\gafs{{{\gaf}^2}}
\def\gvl{{g_\sst{V}^\ell}}
\def\gal{{g_\sst{A}^\ell}}
\def\gve{{g_\sst{V}^e}}
\def\gae{{g_\sst{A}^e}}
\def\gvnu{{g_\sst{V}^\nu}}
\def\ganu{{g_\sst{A}^\nu}}
\def\gvu{{g_\sst{V}^u}}
\def\gau{{g_\sst{A}^u}}
\def\gvd{{g_\sst{V}^d}}
\def\gad{{g_\sst{A}^d}}
\def\gvs{{g_\sst{V}^s}}
\def\gas{{g_\sst{A}^s}}
\def\fa{{F_\sst{A}}}
\def\famq{{F_\sst{A}^{many-quark}}}
\def\faoq{{F_\sst{A}^{one-quark}}}
\def\fahad{{F_\sst{A}^\sst{HAD}}}
\def\fan{{F_\sst{A}^\sst{N}}}
\def\ncf{{N_c^f}}
\def\pol{{\varepsilon}}
\def\polp{{\varepsilon^{\>\prime}}}
\def\pv{{\vec p}}
\def\pvs{{{\vec p}^{\>2}}}
\def\ppv{{{\vec p}^{\>\prime}}}
\def\ppvs{{{\vec p}^{\>\prime\>2}}}
\def\qv{{\vec q}}
\def\qvs{{{\vec q}^{\>2}}}
\def\xv{{\vec x}}
\def\xpv{{{\vec x}^{\>\prime}}}
\def\yv{{\vec y}}
\def\tauv{{\vec\tau}}
\def\sigv{{\vec\sigma}}
\def\gry{{{\overrightarrow\nabla}_y}}
\def\grx{{{\overrightarrow\nabla}_x}}
\def\grxp{{{\overrightarrow\nabla}_{x'}}}
\def\sqr#1#2{{\vcenter{\vbox{\hrule height.#2pt
		\hbox{\vrule width.#2pt height#1pt \kern#1pt
			\vrule width.#2pt}
		\hrule height.#2pt}}}}
\def\square{{\mathchoice\sqr74\sqr74\sqr{6.3}3\sqr{3.5}3}}
\def\arad{{A^{rad}_\sst{PNC}}}
\def\avaffp{{A^{ff'}_\sst{VA}}}
\def\avafpf{{A^{f'f}_\sst{VA}}}
\def\rvaffp{{R^{ff'}_\sst{VA}}}
\def\vpp{{V^{pp'}}}
\def\vkk{{V^{kk'}}}
\def\app{{A^{pp'}}}
\def\akk{{A^{kk'}}}
\def\gmass{{\Gamma_\sst{MASS}}}
\def\hcal#1{{\hbox{\cal #1}}}
\def\sst#1{{\scriptscriptstyle #1}}
\def\Jem{{J_\mu^{em}}}
\def\CALA{{\hbox{\cal A}}}
\def\CALL{{\hbox{\cal L}}}
\def\mpi{{m_\pi}}
\def\mpis{{m^2_\pi}}
\def\notq{{\not\! q}}
\def\notp{{\not\! p}}
\def\notpp{{{\not\! p}^{\,\prime}}}
\def\notk{{\not\! k}}
\def\mn{{m_\sst{N}}}
\def\mns{{m^2_\sst{N}}}
\def\mpibar{{{\overline m}_\pi}}
\def\mnbar{{\overline\mn}}
\def\mk{{m_\sst{K}}}
\def\msig{{m_\sst{\Sigma}}}
\def\mvm{{m_\sst{VM}}}
\def\mvms{{m^2_\sst{VM}}}
\def\mro{{m_\rho}}
\def\mros{{m^2_\rho}}
\def\cvg{{C_{\sst{V}\gamma}}}
\def\crog{{C_{\rho\gamma}}}
\def\dels{{\Delta\hbox{S}}}
\def\gpnn{{g_{\sst{NN}\pi}}}
\def\grnn{{g_{\sst{NN}\rho}}}
\def\gnnm{{g_\sst{NNM}}}
\def\hnnm{{h_\sst{NNM}}}
\def\Gf{{G_\sst{F}}}
\def\subar{{\overline u}}
\def\lws{{\hcal{L}_\sst{WS}^{classical}}}
\def\obs{{\hcal{O}^\sst{PNC}}}
\def\obsatom{{\hcal{O}^\sst{PNC}_{atom}}}
\def\obsnuc{{\hcal{O}^\sst{PNC}_{nuc}}}
\def\Jhat{{\hat J}}
\def\Hhat{{\hat H}}
\def\kn{{\kappa_n}}
\def\kp{{\kappa_p}}
\def\fft{{{\tilde F}_2^{(o)}}}
\def\Rbar{{\bar R}}
\def\Rtil{{\tilde R}}
\def\HPNC{{\Hhat(2)_\sst{PNC}^\sst{NUC}}}
\def\Hweak{{\hcal{H}_\sst{W}^\sst{PNC}}}
\def\rfsem{{\langle R_5^2\rangle_{em}}}
\def\sst#1{{\scriptscriptstyle #1}}
\def\hcal#1{{\hbox{\cal #1}}}
\def\eexp#1{{\hbox{e}^{#1}}}
\def\ahat{{\hat a}}
\def\Jhat{{\hat J}}
\def\Hhat{{\hat H}}
\def\That{{\hat T}}
\def\Chat{{\hat C}}
\def\Ohat{{\hat O}}
\def\Lhat{{\hat L}}
\def\Phat{{\hat P}}
\def\Mhat{{\hat M}}
\def\Shat{{\hat S}}
\def\Rhat{{\hat R}}
\def\rohat{{\hat\rho}}
\def\ehat{{\hat e}}
\def\OP{{\hat\hcal{O}}}
\def\mn{{m_\sst{N}}}
\def\mns{{m_\sst{N}^2}}
\def\mni{{m_\sst{N}^{-1}}}
\def\mnis{{m_\sst{N}^{-2}}}
\def\mnic{{m_\sst{N}^{-3}}}
\def\mpi{{m_\pi}}
\def\mpis{{m^2_\pi}}
\def\cpv{{\vec P}}
\def\cppv{{{\vec P}^{\>\prime}}}
\def\qv{{\vec q}}
\def\pv{{\vec p}}
\def\ppv{{{\vec p}^{\>\prime}}}
\def\kv{{\vec k}}
\def\qvs{{\qv^{\> 2}}}
\def\pvs{{\pv^{\> 2}}}
\def\ppvs{{{\vec p}^{\>\prime\>2}}}
\def\kvs{{kv^{\, 2}}}
\def\xv{{\vec x}}
\def\xpv{{{\vec x}^{\>\prime}}}
\def\yv{{\vec y}}
\def\rv{{\vec r}}
\def\Rv{{\vec R}}
\def\Jv{{\vec J}}
\def\sigv{{\vec\sigma}}
\def\tauv{{\vec\tau}}
\def\Yvh{{\vec Y}}
\def\grad{{\vec\nabla}}
\def\Gf{{G_\sst{F}}}
\def\gpnn{{g_{\pi\sst{NN}}}}
\def\fpi{{f_\pi}}
\def\notk{{\rlap/k}}
\def\notp{{\rlap/p}}
\def\notpp{{{\notp}^{\>\prime}}}
\def\notq{{\rlap/q}}
\def\ubar{{\bar u}}
\def\vbar{{\bar v}}
\def\Nbar{{\overline N}}
\def\rbra#1{{\langle#1\parallel}}
\def\rket#1{{\parallel#1\rangle}}
\def\lpi{{L_\pi}}
\def\lpis{{L_\pi^2}}
\def\gfpi{\frac{\gpnn\fpi}{ 8\sqrt{2}\pi\mn}}
\def\kf{{k_\sst{F}}}
\def\rhoa{{\rho_\sst{A}}}
\def\kt{{\tilde k}}
\def\mpit{{{\tilde m}_\pi}}
\def\mpits{{{\tilde m}_\pi^2}}
\def\jJ{{j_\sst{J}}}
\def\jL{{j_\sst{L}}}
\def\lws{{\hcal{L}_\sst{WS}^{cl}}}
\def\famc{{F_\sst{A}^{meson\, cloud}}}
\def\faob{{F_\sst{A}^{one-body}}}
\def\famb{{F_\sst{A}^{many-body}}}
\def\xpibar{{x_\pi}}

\def\xivz{{\xi_\sst{V}^{(0)}}}
\def\xivt{{\xi_\sst{V}^{(3)}}}
\def\xive{{\xi_\sst{V}^{(8)}}}
\def\xiaz{{\xi_\sst{A}^{(0)}}}
\def\xiat{{\xi_\sst{A}^{(3)}}}
\def\xiae{{\xi_\sst{A}^{(8)}}}
\def\xivtez{{\xi_\sst{V}^{T=0}}}
\def\xivteo{{\xi_\sst{V}^{T=1}}}
\def\xiatez{{\xi_\sst{A}^{T=0}}}
\def\xiateo{{\xi_\sst{A}^{T=1}}}
\def\xiva{{\xi_\sst{V,A}}}

\def\rvz{{R_\sst{V}^{(0)}}}
\def\rvt{{R_\sst{V}^{(3)}}}
\def\rve{{R_\sst{V}^{(8)}}}
\def\raz{{R_\sst{A}^{(0)}}}
\def\rat{{R_\sst{A}^{(3)}}}
\def\rae{{R_\sst{A}^{(8)}}}
\def\rvtez{{R_\sst{V}^{T=0}}}
\def\rvteo{{R_\sst{V}^{T=1}}}
\def\ratez{{R_\sst{A}^{T=0}}}
\def\rateo{{R_\sst{A}^{T=1}}}

\def\mapright#1{\smash{\mathop{\longrightarrow}\limits^{#1}}}

\def\FOS{{F_1^{(s)}}}
\def\FTS{{F_2^{(s)}}}
\def\GAS{{G_\sst{A}^{(s)}}}
\def\GES{{G_\sst{E}^{(s)}}}
\def\GMS{{G_\sst{M}^{(s)}}}
\def\GATEZ{{G_\sst{A}^{\sst{T}=0}}}
\def\GATEO{{G_\sst{A}^{\sst{T}=1}}}
\def\mdax{{M_\sst{A}}}
\def\mustr{{\mu_s}}
\def\rsstr{{r^2_s}}
\def\rhostr{{\rho_s}}
\def\GEG{{G_\sst{E}^\gamma}}
\def\GEZ{{G_\sst{E}^\sst{Z}}}
\def\GMG{{G_\sst{M}^\gamma}}
\def\GMZ{{G_\sst{M}^\sst{Z}}}
\def\GEn{{G_\sst{E}^n}}
\def\GEp{{G_\sst{E}^p}}
\def\GMn{{G_\sst{M}^n}}
\def\GMp{{G_\sst{M}^p}}
\def\GAp{{G_\sst{A}^p}}
\def\GAn{{G_\sst{A}^n}}
\def\GA{{G_\sst{A}}}
\def\GETEZ{{G_\sst{E}^{\sst{T}=0}}}
\def\GETEO{{G_\sst{E}^{\sst{T}=1}}}
\def\GMTEZ{{G_\sst{M}^{\sst{T}=0}}}
\def\GMTEO{{G_\sst{M}^{\sst{T}=1}}}
\def\lamd{{\lambda_\sst{D}^\sst{V}}}
\def\lamn{{\lambda_n}}
\def\lams{{\lambda_\sst{E}^{(s)}}}
\def\bvz{{\beta_\sst{V}^0}}
\def\bvo{{\beta_\sst{V}^1}}
\def\Gdip{{G_\sst{D}^\sst{V}}}
\def\GdipA{{G_\sst{D}^\sst{A}}}

\def\RAp{{R_\sst{A}^p}}
\def\RAn{{R_\sst{A}^n}}
\def\RVp{{R_\sst{V}^p}}
\def\RVn{{R_\sst{V}^n}}
\def\rva{{R_\sst{V,A}}}

\def\jnc{{J^\sst{NC}_\mu}}
\def\jncf{{J^\sst{NC}_{\mu 5}}}
\def\jem{{J^\sst{EM}_\mu}}
\def\ftil{{\tilde F}}
\def\ftilo{{\tilde F_1}}
\def\ftilt{{\tilde F_2}}
\def\gtil{{\tilde G}}
\def\gtila{{\tilde G_\sst{A}}}
\def\gtilp{{\tilde G_\sst{P}}}
\def\geptil{{\tilde G_\sst{E}^p}}
\def\gmptil{{\tilde G_\sst{M}^p}}
\def\gentil{{\tilde G_\sst{E}^n}}
\def\gmntil{{\tilde G_\sst{M}^n}}
\def\geteztil{{{\tilde G}_\sst{E}^{\sst{T}=0}}}
\def\gmteztil{{{\tilde G}_\sst{M}^{\sst{T}=0}}}
\def\geteotil{{{\tilde G}_\sst{E}^{\sst{T}=1}}}
\def\gmteztil{{{\tilde G}_\sst{M}^{\sst{T}=1}}}

\def\vL{{v_\sst{L}}}
\def\vT{{v_\sst{T}}}
\def\vTp{{v_\sst{T'}}}
\def\RL{{R_\sst{L}}}
\def\RT{{R_\sst{T}}}
\def\WAVL{{W_\sst{AV}^\sst{L}}}
\def\WAVT{{W_\sst{AV}^\sst{T}}}
\def\WVATp{{W_\sst{VA}^\sst{T'}}}

\def\bra#1{{\langle#1\vert}}
\def\ket#1{{\vert#1\rangle}}
\def\coeff#1#2{{\scriptstyle\frac{#1}{#2}}}
\def\undertext#1{{$\underline{\hbox{#1}}$}}
\def\hcal#1{{\hbox{\cal #1}}}
\def\sst#1{{\scriptscriptstyle #1}}
\def\eexp#1{{\hbox{e}^{#1}}}
\def\rbra#1{{\langle #1 \vert\!\vert}}
\def\rket#1{{\vert\!\vert #1\rangle}}
\def\lsim{{ <\atop\sim}}
\def\gsim{{ >\atop\sim}}
\def\refmark#1{{$^{\hbox{#1}}$}}
\def\nubar{{\bar\nu}}
\def\Gmu{{G_\mu}}
\def\alr{{A_\sst{LR}}}
\def\wpv{{W^\sst{PV}}}
\def\evec{{\vec e}}
\def\notq{{\not\! q}}
\def\notk{{\not\! k}}
\def\notp{{\not\! p}}
\def\notpp{{\not\! p'}}
\def\notder{{\not\! \partial}}
\def\notcder{{\not\!\! D}}
\def\Jem{{J_\mu^{em}}}
\def\Jana{{J_{\mu 5}^{anapole}}}
\def\nue{{\nu_e}}
\def\mn{{m_\sst{N}}}
\def\mns{{m^2_\sst{N}}}
\def\me{{m_e}}
\def\mes{{m^2_e}}
\def\mmu{{m_\mu}}
\def\mmus{{m^2_\mu}}
\def\mf{{m_f}}
\def\mfs{{m_f^2}}
\def\mfp{{m_{f'}}}
\def\mfps{{m_{f'}^2}}
\def\mq{{m_q}}
\def\mqs{{m_q^2}}
\def\ml{{m_\ell}}
\def\mls{{m_\ell^2}}
\def\mt{{m_t}}
\def\mts{{m_t^2}}
\def\mnu{{m_\nu}}
\def\mnus{{m_\nu^2}}
\def\mz{{M_\sst{Z}}}
\def\mzs{{M^2_\sst{Z}}}
\def\mw{{M_\sst{W}}}
\def\mws{{M^2_\sst{W}}}
\def\mh{{M_\sst{H}}}
\def\mhs{{M^2_\sst{H}}}
\def\ubar{{\bar u}}
\def\dbar{{\bar d}}
\def\sbar{{\bar s}}
\def\qbar{{\bar q}}
\def\Abar{{\overline A}}
\def\Nbar{{\overline N}}
\def\ucr{{u^{\dag}}}
\def\dcr{{d^{\dag}}}
\def\QM{{\sst{QM}}}
\def\ctw{{\cos\theta_\sst{W}}}
\def\stw{{\sin\theta_\sst{W}}}
\def\sstw{{\sin^2\theta_\sst{W}}}
\def\cstw{{\cos^2\theta_\sst{W}}}
\def\cftw{{\cos^4\theta_\sst{W}}}
\def\tw{{\theta_\sst{W}}}
\def\sstwh{{\sin^2{\hat\theta}_\sst{W}}}
\def\sstwb{{\sin^2{\bar\theta}_\sst{W}}}
\def\ztil{{{\tilde Z}^{1/2}}}
\def\ztilij{{{\tilde Z}^{1/2}_{ij}}}
\def\zstil{{\tilde Z}}
\def\zren{{Z^{1/2}}}
\def\zrenw{{Z^{1/2}_\sst{WW}}}
\def\zrenz{{Z^{1/2}_\sst{ZZ}}}
\def\zrena{{Z^{1/2}_\sst{AA}}}
\def\zrenaz{{Z^{1/2}_\sst{AZ}}}
\def\zrenza{{Z^{1/2}_\sst{ZA}}}
\def\zrenl{{Z^{1/2}_\sst{L}}}
\def\zrenr{{Z^{1/2}_\sst{R}}}
\def\zrenps{{Z^{1/2}_\psi}}
\def\zrenpsb{{Z^{1/2}_{\bar\psi}}}
\def\znren{{Z^{-1/2}}}
\def\dw{{\delta_\sst{W}}}
\def\dz{{\delta_\sst{Z}}}
\def\dzb{{{\overline\delta}_\sst{Z}}}
\def\da{{\delta_\sst{A}}}
\def\dza{{\delta_\sst{ZA}}}
\def\dzap{{\delta^{pole}_\sst{ZA}}}
\def\dzab{{{\overline\delta}_\sst{ZA}}}
\def\daz{{\delta_\sst{AZ}}}
\def\dazp{{\delta^{pole}_\sst{AZ}}}
\def\dazb{{{\overline\delta}_\sst{AZ}}}
\def\dmw{{\delta M^2_\sst{W}}}
\def\dmz{{\delta M^2_\sst{Z}}}
\def\dmwb{{{\overline\dmw}}}
\def\dmzb{{{\overline\dmz}}}
\def\dy{{\delta_\sst{Y}}}
\def\dyb{{{\overline\delta}_\sst{Y}}}
\def\dps{{\delta_\psi}}
\def\dpsf{{\delta_\psi^5}}
\def\dl{{\delta_\sst{L}}}
\def\dr{{\delta_\sst{R}}}
\def\tmunu{{T_{\mu\nu}}}
\def\lmunu{{L_{\mu\nu}}}
\def\gp{{(\xi-1)}}
\def\cc{\frac{\alpha}{(4\pi)}}
\def\auv{{\alpha_\sst{UV}}}
\def\air{{\alpha_\sst{IR}}}
\def\qw{{Q_\sst{W}^2}}
\def\Gf{{G_\sst{F}}}
\def\gv{{g_\sst{V}}}
\def\ga{{g_\sst{A}}}
\def\gvq{{g_\sst{V}^q}}
\def\gaq{{g_\sst{A}^q}}
\def\gvf{{g_\sst{V}^{f}}}
\def\gaf{{g_\sst{A}^{f}}}
\def\gvfp{{g_\sst{V}^{f'}}}
\def\gafp{{g_\sst{A}^{f'}}}
\def\gvfs{{{\gvf}^2}}
\def\gafs{{{\gaf}^2}}
\def\gvl{{g_\sst{V}^\ell}}
\def\gal{{g_\sst{A}^\ell}}
\def\gve{{g_\sst{V}^e}}
\def\gae{{g_\sst{A}^e}}
\def\gvnu{{g_\sst{V}^\nu}}
\def\ganu{{g_\sst{A}^\nu}}
\def\gvu{{g_\sst{V}^u}}
\def\gau{{g_\sst{A}^u}}
\def\gvd{{g_\sst{V}^d}}
\def\gad{{g_\sst{A}^d}}
\def\gvs{{g_\sst{V}^s}}
\def\gas{{g_\sst{A}^s}}
\def\fa{{F_\sst{A}}}
\def\famq{{F_\sst{A}^{many-quark}}}
\def\faoq{{F_\sst{A}^{one-quark}}}
\def\fahad{{F_\sst{A}^\sst{HAD}}}
\def\fan{{F_\sst{A}^\sst{N}}}
\def\ncf{{N_c^f}}
\def\pol{{\varepsilon}}
\def\polp{{\varepsilon^{\>\prime}}}
\def\pv{{\vec p}}
\def\pvs{{{\vec p}^{\>2}}}
\def\ppv{{{\vec p}^{\>\prime}}}
\def\ppvs{{{\vec p}^{\>\prime\>2}}}
\def\qv{{\vec q}}
\def\qvs{{{\vec q}^{\>2}}}
\def\xv{{\vec x}}
\def\xpv{{{\vec x}^{\>\prime}}}
\def\yv{{\vec y}}
\def\tauv{{\vec\tau}}
\def\sigv{{\vec\sigma}}
\def\gry{{{\overrightarrow\nabla}_y}}
\def\grx{{{\overrightarrow\nabla}_x}}
\def\grxp{{{\overrightarrow\nabla}_{x'}}}
\def\sqr#1#2{{\vcenter{\vbox{\hrule height.#2pt
		\hbox{\vrule width.#2pt height#1pt \kern#1pt
			\vrule width.#2pt}
		\hrule height.#2pt}}}}
\def\square{{\mathchoice\sqr74\sqr74\sqr{6.3}3\sqr{3.5}3}}
\def\arad{{A^{rad}_\sst{PNC}}}
\def\avaffp{{A^{ff'}_\sst{VA}}}
\def\avafpf{{A^{f'f}_\sst{VA}}}
\def\rvaffp{{R^{ff'}_\sst{VA}}}
\def\vpp{{V^{pp'}}}
\def\vkk{{V^{kk'}}}
\def\app{{A^{pp'}}}
\def\akk{{A^{kk'}}}
\def\gmass{{\Gamma_\sst{MASS}}}

\def\hcal#1{{\hbox{\cal #1}}}
\def\sst#1{{\scriptscriptstyle #1}}
\def\Jem{{J_\mu^{em}}}
\def\CALA{{\hbox{\cal A}}}
\def\CALL{{\hbox{\cal L}}}
\def\mpi{{m_\pi}}
\def\mpis{{m^2_\pi}}
\def\notq{{\not\! q}}
\def\notp{{\not\! p}}
\def\notpp{{{\not\! p}^{\,\prime}}}
\def\notk{{\not\! k}}
\def\mn{{m_\sst{N}}}
\def\mns{{m^2_\sst{N}}}
\def\mpibar{{{\overline m}_\pi}}
\def\mnbar{{\overline\mn}}
\def\mk{{m_\sst{K}}}
\def\msig{{m_\sst{\Sigma}}}
\def\mvm{{m_\sst{VM}}}
\def\mvms{{m^2_\sst{VM}}}
\def\mro{{m_\rho}}
\def\mros{{m^2_\rho}}
\def\cvg{{C_{\sst{V}\gamma}}}
\def\crog{{C_{\rho\gamma}}}
\def\dels{{\Delta\hbox{S}}}
\def\gpnn{{g_{\sst{NN}\pi}}}
\def\grnn{{g_{\sst{NN}\rho}}}
\def\gnnm{{g_\sst{NNM}}}
\def\hnnm{{h_\sst{NNM}}}
\def\Gf{{G_\sst{F}}}
\def\subar{{\overline u}}
\def\lws{{\hcal{L}_\sst{WS}^{classical}}}
\def\obs{{\hcal{O}^\sst{PNC}}}
\def\obsatom{{\hcal{O}^\sst{PNC}_{atom}}}
\def\obsnuc{{\hcal{O}^\sst{PNC}_{nuc}}}
\def\Jhat{{\hat J}}
\def\Hhat{{\hat H}}
\def\kn{{\kappa_n}}
\def\kp{{\kappa_p}}
\def\fft{{{\tilde F}_2^{(o)}}}
\def\Rbar{{\bar R}}
\def\Rtil{{\tilde R}}
\def\HPNC{{\Hhat(2)_\sst{PNC}^\sst{NUC}}}
\def\Hweak{{\hcal{H}_\sst{W}^\sst{PNC}}}
\def\rfsem{{\langle R_5^2\rangle_{em}}}
\def\sst#1{{\scriptscriptstyle #1}}
\def\hcal#1{{\hbox{\cal #1}}}
\def\eexp#1{{\hbox{e}^{#1}}}
\def\ahat{{\hat a}}
\def\Jhat{{\hat J}}
\def\Hhat{{\hat H}}
\def\That{{\hat T}}
\def\Chat{{\hat C}}
\def\Ohat{{\hat O}}
\def\Lhat{{\hat L}}
\def\Phat{{\hat P}}
\def\Mhat{{\hat M}}
\def\Shat{{\hat S}}
\def\Rhat{{\hat R}}
\def\rohat{{\hat\rho}}
\def\ehat{{\hat e}}
\def\OP{{\hat\hcal{O}}}
\def\mn{{m_\sst{N}}}
\def\mns{{m_\sst{N}^2}}
\def\mni{{m_\sst{N}^{-1}}}
\def\mnis{{m_\sst{N}^{-2}}}
\def\mnic{{m_\sst{N}^{-3}}}
\def\mpi{{m_\pi}}
\def\mpis{{m^2_\pi}}
\def\cpv{{\vec P}}
\def\cppv{{{\vec P}^{\>\prime}}}
\def\qv{{\vec q}}
\def\pv{{\vec p}}
\def\ppv{{{\vec p}^{\>\prime}}}
\def\kv{{\vec k}}
\def\qvs{{\qv^{\> 2}}}
\def\pvs{{\pv^{\> 2}}}
\def\ppvs{{{\vec p}^{\>\prime\>2}}}
\def\kvs{{kv^{\, 2}}}
\def\xv{{\vec x}}
\def\xpv{{{\vec x}^{\>\prime}}}
\def\yv{{\vec y}}
\def\rv{{\vec r}}
\def\Rv{{\vec R}}
\def\Jv{{\vec J}}
\def\sigv{{\vec\sigma}}
\def\tauv{{\vec\tau}}
\def\Yvh{{\vec Y}}
\def\grad{{\vec\nabla}}
\def\Gf{{G_\sst{F}}}
\def\gpnn{{g_{\pi\sst{NN}}}}
\def\fpi{{f_\pi}}
\def\notk{{\rlap/k}}
\def\notp{{\rlap/p}}
\def\notpp{{{\notp}^{\>\prime}}}
\def\notq{{\rlap/q}}
\def\ubar{{\bar u}}
\def\vbar{{\bar v}}
\def\Nbar{{\over
line N}}
\def\rbra#1{{\langle#1\parallel}}
\def\rket#1{{\parallel#1\rangle}}
\def\lpi{{L_\pi}}
\def\lpis{{L_\pi^2}}
\def\kf{{k_\sst{F}}}
\def\rhoa{{\rho_\sst{A}}}
\def\kt{{\tilde k}}
\def\mpit{{{\tilde m}_\pi}}
\def\mpits{{{\tilde m}_\pi^2}}
\def\jJ{{j_\sst{J}}}
\def\jL{{j_\sst{L}}}
\def\lws{{\hcal{L}_\sst{WS}^{cl}}}
\def\famc{{F_\sst{A}^{meson\, cloud}}}
\def\faob{{F_\sst{A}^{one-body}}}
\def\famb{{F_\sst{A}^{many-body}}}
\def\xpibar{{x_\pi}}

\def\xivz{{\xi_\sst{V}^{(0)}}}
\def\xivt{{\xi_\sst{V}^{(3)}}}
\def\xive{{\xi_\sst{V}^{(8)}}}
\def\xiaz{{\xi_\sst{A}^{(0)}}}
\def\xiat{{\xi_\sst{A}^{(3)}}}
\def\xiae{{\xi_\sst{A}^{(8)}}}
\def\xivtez{{\xi_\sst{V}^{T=0}}}
\def\xivteo{{\xi_\sst{V}^{T=1}}}
\def\xiatez{{\xi_\sst{A}^{T=0}}}
\def\xiateo{{\xi_\sst{A}^{T=1}}}
\def\xiva{{\xi_\sst{V,A}}}

\def\rvz{{R_\sst{V}^{(0)}}}
\def\rvt{{R_\sst{V}^{(3)}}}
\def\rve{{R_\sst{V}^{(8)}}}
\def\raz{{R_\sst{A}^{(0)}}}
\def\rat{{R_\sst{A}^{(3)}}}
\def\rae{{R_\sst{A}^{(8)}}}
\def\rvtez{{R_\sst{V}^{T=0}}}
\def\rvteo{{R_\sst{V}^{T=1}}}
\def\ratez{{R_\sst{A}^{T=0}}}
\def\rateo{{R_\sst{A}^{T=1}}}

\def\mapright#1{\smash{\mathop{\longrightarrow}\limits^{#1}}}

\def\FOS{{F_1^{(s)}}}
\def\FTS{{F_2^{(s)}}}
\def\GAS{{G_\sst{A}^{(s)}}}
\def\GES{{G_\sst{E}^{(s)}}}
\def\GMS{{G_\sst{M}^{(s)}}}
\def\GATEZ{{G_\sst{A}^{\sst{T}=0}}}
\def\GATEO{{G_\sst{A}^{\sst{T}=1}}}
\def\mdax{{M_\sst{A}}}
\def\mustr{{\mu_s}}
\def\rsstr{{r^2_s}}
\def\rhostr{{\rho_s}}
\def\GEG{{G_\sst{E}^\gamma}}
\def\GEZ{{G_\sst{E}^\sst{Z}}}
\def\GMG{{G_\sst{M}^\gamma}}
\def\GMZ{{G_\sst{M}^\sst{Z}}}
\def\GEn{{G_\sst{E}^n}}
\def\GEp{{G_\sst{E}^p}}
\def\GMn{{G_\sst{M}^n}}
\def\GMp{{G_\sst{M}^p}}
\def\GAp{{G_\sst{A}^p}}
\def\GAn{{G_\sst{A}^n}}
\def\GA{{G_\sst{A}}}
\def\GETEZ{{G_\sst{E}^{\sst{T}=0}}}
\def\GETEO{{G_\sst{E}^{\sst{T}=1}}}
\def\GMTEZ{{G_\sst{M}^{\sst{T}=0}}}
\def\GMTEO{{G_\sst{M}^{\sst{T}=1}}}
\def\lamd{{\lambda_\sst{D}^\sst{V}}}
\def\lamn{{\lambda_n}}
\def\lams{{\lambda_\sst{E}^{(s)}}}
\def\bvz{{\beta_\sst{V}^0}}
\def\bvo{{\beta_\sst{V}^1}}
\def\Gdip{{G_\sst{D}^\sst{V}}}
\def\GdipA{{G_\sst{D}^\sst{A}}}

\def\RAp{{R_\sst{A}^p}}
\def\RAn{{R_\sst{A}^n}}
\def\RVp{{R_\sst{V}^p}}
\def\RVn{{R_\sst{V}^n}}
\def\rva{{R_\sst{V,A}}}

\def\jnc{{J^\sst{NC}_\mu}}
\def\jncf{{J^\sst{NC}_{\mu 5}}}
\def\jem{{J^\sst{EM}_\mu}}
\def\ftil{{\tilde F}}
\def\ftilo{{\tilde F_1}}
\def\ftilt{{\tilde F_2}}
\def\gtil{{\tilde G}}
\def\gtila{{\tilde G_\sst{A}}}
\def\gtilp{{\tilde G_\sst{P}}}
\def\geptil{{\tilde G_\sst{E}^p}}
\def\gmptil{{\tilde G_\sst{M}^p}}
\def\gentil{{\tilde G_\sst{E}^n}}
\def\gmntil{{\tilde G_\sst{M}^n}}
\def\geteztil{{{\tilde G}_\sst{E}^{\sst{T}=0}}}
\def\gmteztil{{{\tilde G}_\sst{M}^{\sst{T}=0}}}
\def\geteotil{{{\tilde G}_\sst{E}^{\sst{T}=1}}}
\def\gmteztil{{{\tilde G}_\sst{M}^{\sst{T}=1}}}

\def\vL{{v_\sst{L}}}
\def\vT{{v_\sst{T}}}
\def\vTp{{v_\sst{T'}}}
\def\RL{{R_\sst{L}}}
\def\RT{{R_\sst{T}}}
\def\WAVL{{W_\sst{AV}^\sst{L}}}
\def\WAVT{{W_\sst{AV}^\sst{T}}}
\def\WVATp{{W_\sst{VA}^\sst{T'}}}

\def\rbra#1{{\langle#1\parallel}}
\def\rket#1{{\parallel#1\rangle}}
\def\famc{{F_\sst{A}^{meson\, cloud}}}
\def\faob{{F_\sst{A}^{one-body}}}
\def\dellr{{\Delta_\sst{LR}}}
\def\evec{{\vec e}}
\def\famb{{F_\sst{A}^{many-body}}}
\def\mustr{{\mu_s}}
\def\muiso{{\mu^\sst{T=0}}}
\def\mupro{{\mu^\sst{P}}}
\def\muneu{{\mu^\sst{N}}}
\def\gastr{{G_\sst{A}^{(s)}}}
\def\md{{M_\sst{D}}}
\def\rstr{{\langle r^2\rangle^{(s)}}}
\def\geiso{{G_\sst{E}^\sst{T=0}}}
\def\gmiso{{G_\sst{M}^\sst{T=0}}}
\def\nudvert{{\vert_{\nu D}}}
\def\nubardvert{{\vert_{{\bar\nu}D}}}
\def\rone{{R_{(1)}}}
\def\rtwo{{R_{(2)}}}
\def\rthree{{R_{(3)}}}
\def\rfour{{R_{(4)}}}
\def\feostr{{{\tilde F}_\sst{E1}^{(s)}}}
\def\delnnb{{\Delta_{\nu\bar\nu}}}
\def\xivisos{{\sqrt{3}[\xiv^{(8)}+\coeff{2}{\sqrt{3}}\xiv^{(0)}]}}
\def\xiaisos{{\sqrt{3}[\xia^{(8)}+\coeff{2}{\sqrt{3}}\xia^{(0)}]}}
\def\rvisos{{R_\sst{V}^\sst{T=0}}}
\def\raisos{{R_\sst{A}^\sst{T=0}}}
\def\rvo{{R_\sst{V}^{(0)}}}
\def\rao{{R_\sst{A}^{(0)}}}
\def\wpnc{{W^\sst{PNC}}}
\def\delone{{\Delta_{(1)}}}
\def\deltwo{{\Delta_{(2)}}}
\def\delthree{{\Delta_{(3)}}}
\def\deltwom{{\Delta_{(2)}^\sst{M}}}
\def\deltwoe{{\Delta_{(2)}^\sst{E}}}
\def\feone{{F_\sst{E1}^5}}
\def\rviso{{R_v^\sst{T=0}}}
\def\xia{{\xi_\sst{A}}}
\def\xiv{{\xi_\sst{V}}}
\def\fcz{{F_\sst{C0}}}
\def\fct{{F_\sst{C2}}}
\def\fmo{{F_\sst{M1}}}
\def\dc{{D_\sst{C}}}
\def\dq{{D_\sst{Q}}}
\def\daa{{D_\sst{A}}}
\def\dmm{{D_\sst{M}^\sst{M}}}
\def\dme{{D_\sst{M}^\sst{E}}}
\def\GA{{G_\sst{A}}}
\def\GAstr{{G_\sst{A}^{(s)}}}
\def\GApro{{G_\sst{A}^\sst{P}}}
\def\GAiso{{G_\sst{A}^\sst{T=0}}}
\def\GAisov{{G_\sst{A}^\sst{T=1}}}
\def\fcjstr{{{\tilde F}_\sst{CJ}^{(s)}}}
\def\fmostr{{{\tilde F}_\sst{M1}^{(s)}}}
\def\gestr{{{\tilde G}_\sst{E}^{(s)}}}
\def\gmstr{{{\tilde G}_\sst{M}^{(s)}}}
\def\gmpro{{G_\sst{M}^\sst{P}}}
\def\fmoiso{{F_\sst{M1}^\sst{T=0}}}
\def\delnuc{{\delta_\sst{NUC}}}
\def\nubar{{\bar\nu}}
\def\qw{{Q_\sst{W}}}
\def\alrc{{\alr(^{12}{\rm C})}}
\def\qwcs{{\qw(^{133}{\rm Cs})}}
\def\PR#1{{\it Phys. Rev.} {\bf #1} }
\def\PRC#1{{\it Phys. Rev.} {\bf C#1} }
\def\PRD#1{{\it Phys. Rev.} {\bf D#1} }
\def\PRL#1{{\it Phys. Rev. Lett.} {\bf #1} }
\def\NPA#1{{\it Nucl. Phys.} {\bf A#1} }
\def\NPB#1{{\it Nucl. Phys.} {\bf B#1} }
\def\AoP#1{{\it Ann. of Phys.} {\bf #1} }
\def\PRp#1{{\it Phys. Reports} {\bf #1} }
\def\PLB#1{{\it Phys. Lett.} {\bf B#1} }
\def\ZPA#1{{\it Z. f\"ur Phys.} {\bf A#1} }
\def\ZPC#1{{\it Z. f\"ur Phys.} {\bf C#1} }
\def\etal{{\it et al.}}
\title{  A Strange Mesonic Transition Form Factor \footnote{This
work is supported in part by funds
provided by the U. S. Department of Energy (D.O.E.) under contracts
\#DE-AC05-84ER40150 and \#DE-FG05-94ER0832.}}
\author{J. L. Goity}
\address{  Department of Physics, Hampton University,
 Hampton, VA 23668, and \\ Theory Group, MS 12H2,
  Continuous Electron Beam Accelerator Facility,\\
  Newport News, VA  23606, U.S.A.}

\author{M. J. Musolf \footnote{National Science Foundation
Young Investigator}}
\address{  Department of Physics, Old Dominion University
 Norfolk, VA 23529, and \\
 Theory Group, MS 12H2
  Continuous Electron Beam Accelerator Facility,\\
  Newport News, VA  23606, U.S.A.}
\maketitle
\newpage
\begin{abstract}
{\small {The strange-quark vector current $\rho$-to-$\pi$ meson transition form
factor is computed at one-loop order using strange meson intermediate
states. A comparison is made with a $\phi$-meson dominance model
estimate. We find that one-loop  contributions  are comparable in
magnitude to
those predicted by $\phi$-meson dominance. It is possible
that the one-loop
contribution  can make the matrix element as large as those of the
electromagnetic current
mediating  vector meson radiative decays. However, due to the quadratic
dependence of the  one-loop results on the hadronic form factor cut-off mass, a
large uncertainty
in the  estimate of the loops is unavoidable.
These results indicate that non-nucleonic
strange quarks could contribute appreciably in moderate-$|Q^2|$
parity-violating electron-nucleus scattering measurements aimed at
probing the strange-quark content of the nucleon}}.
\end{abstract}
\newpage

\parskip 3mm
\section{  Introduction} There has been considerable interest recently
in the use of semi-leptonic weak neutral current scattering to study the
strange-quark \lq\lq content" of the nucleon [1-12].
In particular, several
parity-violating (PV) electron scattering experiments are planned and/or
underway at MIT-Bates, CEBAF, and MAINZ whose objective is to measure the
nucleon's strange-quark vector current form factors [13-17]. In a
similar vein,
a low-$|Q^2|$ determination of the nucleon's axial vector strangeness form
factor will be made using neutrino scattering at
LAMPF $\cite{LAMPF1}$, following on the higher-$|Q^2|$
measurement
made at Brookhaven $\cite{ahrens}$.
These experiments are of interest in part because
they provide a new window on the role played by non-valence degrees of freedom
(specifically, virtual $s\sbar$ pairs) in the nucleon's response to a low- or
medium-energy external probe. In contrast to the theoretical analysis of
scattering in the deep inelastic regime,  for which perturbative methods are
applicable, the interpretation of low-to-medium energy scattering
can be carried out at present only within the context of effective hadronic
models. In the case of the nucleon's strangeness form factors, several model
calculations have been performed yielding a rather broad spectrum of
results [20-26].
It is desirable, then, that experimental determinations of these form factors
be carried out at a level of precision allowing one to distinguish among
various models and the physical pictures on which they are based.


	As discussed elsewhere in the literature, semi-leptonic measurements
performed with proton targets alone would not be sufficient for this
purpose $\cite{mike1,mike2,mike3}$.
A program of measurements which includes $A>1$ targets
appears to be warranted $\cite{mike1,mike2}$.
The interpretation of neutrino-{\it nucleus} and
PV electron-{\it nucleus} scattering observables naturally introduces a new
level of complication associated with many-body nuclear dynamics not
encountered in the case of proton targets. These many-body effects are
interesting in two respects. On the one hand, if one wishes to extract the
single nucleon strangeness form factors from nuclear form factors, one
requires knowledge of the many-body contributions to the nuclear strangeness
form factors. On the other, the role played by non-nucleonic strangeness
({\it i.e.}, non-nucleonic, non-valence quark degrees of freedom) in the
{\it nuclear} response is interesting in its own right.

	Recently, one class of many-body contributions to the nuclear
strangeness form factors -- meson exchange currents (MEC's) -- were analyzed
for the case of $^4$He $\cite{mike5,mike6}$.
A helium target will be used in a future
CEBAF PV electron scattering experiment designed to study the nucleon's
strangeness electric form factor at moderate-$|Q^2|$ $\cite{cebaf1}$.
In Ref. $\cite{mike6}$,
it was shown that MEC's give a non-negligible contribution to the $^4$He
strangeness form factor at the kinematics of the approved experiment. Of
particular note is the $\rho$-to-$\pi$-meson strange-quark \lq\lq transition
current". Using a simple $\phi$-meson vector dominance model and the known
$\phi\to\rho\pi$ branching ratio, this MEC was estimated to give a 15\%
contribution to the PV asymmetry. Given the magnitude of this result and
its potential importance in the interpretation of the $^4$He experiment,
one would like a more detailed analysis of this strangeness transition
current. The goal of the present paper is to improve upon the vector
dominance estimate of Refs. $\cite{mike5,mike6}$ by including loop effects.
In so doing,
our objective is not so much to provide an airtight theoretical prediction
as to arrive at an order of magnitude for, and quantify the theoretical
uncertainty associated with, loop effects. Section II gives a brief
discussion of the nuclear physics context for our calculation as well as a
review of the $\phi$-dominance estimate of the $\rho$-$\pi$ strangeness
form factor. In Section III we present the formal framework in which
we carry out our analysis. The loop calculation is presented in Section IV.
Section V gives our results and
a discussion of their significance, and in  Section VI we summarize our work.
Technical details may be found in the Appendix.

\section{ Nuclear Strangeness Form Factors}

\def\fcnucs{{F_\sst{C}^{(s)}}}
\def\fcnuciso{{F_\sst{C}^\sst{T=0}}}

\medskip
The $^4$He PV asymmetry, $\alr$, can be written as $\cite{mike1,mike2,mike3}$
\begin{equation}
\alr=\frac{G_\mu|Q^2|}{
4\,\sqrt{2}\,\pi\,\alpha}\left[4\;\sstw+\frac{\fcnucs(q)}{
\fcnuciso(q)}\right] ,
\end{equation}
where $G_\mu$ is the Fermi constant measured in $\mu$-decay, $Q^2=\omega^2
-|\qv|^2$ is the four-momentum transfer squared, and where electroweak
radiative corrections have been ignored for simplicity. The quantity
$\fcnuciso(q)$ is the $^4$He electromagnetic elastic charge form factor
($T=0$ for isoscalar targets). The $^4$He elastic strangeness form factor,
$\fcnucs(q)$, is given by
\begin{equation}
\fcnucs(q)=\int\ d^3x e^{i\qv\cdot\xv}\;\bra{\rm g.s.} s^{\dag}(\xv) s(\xv)
\ket{\rm g.s.},
\end{equation}
where $\ket{\rm g.s.}$ is the nuclear ground state and $s(\xv)$ is the
strange-quark field operator. Note that since $s^{\dag}s$ is just the
charge component of the strangeness vector current, $\sbar\gamma_\mu s$, and
since $^4$He has no net strangeness, the form factor $\fcnucs$ must vanish
at zero momentum transfer. For non-zero momentum transfer, $\fcnucs(q)$
receives a number of contributions, some of which are illustrated in Fig. 1.

	In conventional (non-relativistic) nuclear models, the processes of
Figs. 1a and 1b are sensitive to the nucleon's strangeness vector current
form factors. Those of type 1a give the so-called impulse approximation, or
one-body, contribution. Processes of type 1b which involve an exchange of
a meson $M$ must be included in any nuclear calculation which respects
vector current conservation and in which the two-nucleon potential arises from
the exchange of meson $M$. In the present paper, we are concerned with
contributions of type 1c, which involve matrix elements of the strangeness
vector current between meson states $M$ and $M'$. It is straightforward to
show using G-parity invariance that $\bra{M'}\sbar\gamma_\mu s\ket{M}$
vanishes when $M=M'$. Hence, only transition matrix elements contribute.
The corresponding MEC's are referred to as \lq\lq transition currents".
For moderate values of momentum-transfer, one expects the lightest mesons
to give the largest effect, since they have the longest range and experience
the least suppression from the $N-N$ short range repulsion. For this reason,
we restrict our attention to the transition current involving the lightest
possible allowed pair of states: $M=\pi$ and $M'=\rho$.

\def\gropisff{{g_{\rho\pi}^{(s)}}}
\def\gropis{{g_{\rho\pi s}}}

	For on-shell mesons, the $\rho$-to-$\pi$ transition matrix element has
the structure
\begin{equation}
\bra{\rho^a(k_1,\varepsilon)}\sbar\gamma_\mu s\ket{\pi^b(k_2)} =
\frac{\gropisff(Q^2)}{M_\rho}\epsilon_{\mu\nu\alpha\beta}k_1^\nu k_2^\alpha
\varepsilon^{\beta\ast}\delta^{ab},
\end{equation}
where $\varepsilon^\beta$ is the $\rho$-meson polarization vector, $k_1$ and
$k_2$ are the meson momenta, and $a$ and $b$ are isospin indices. In the case
of nuclear processes, where the typical momenta of hadrons inside the
nucleus have magnitudes less than the Fermi momentum ($\sim 200$ MeV), the
virtual $\rho$-meson will be rather far off its mass shell. Consequently,
the  dimensionless form factor $\gropisff$ ought also to depend on $k_1^2$ and
$k_2^2$
as well as $Q^2$. It is conventional in nuclear calculations, however, to
neglect the off-shell dependence of transition form factors, so for purposes
of making contact with this framework, we will quote results for the
on-shell case.

	In previous work $\cite{mike5,mike6}$, an estimate of
$\gropisff(Q^2)$ was made
based on the assumption of $\phi$-meson dominance, as illustrated in Fig. 2.
Under this assumption, one has
\begin{equation}
\bra{\rho^a(k_1,\varepsilon)}\sbar\gamma_\mu s \ket{\pi^b(k_2)} =
\bra{0}\sbar\gamma_\mu s\ket{\phi(Q, \varepsilon_\phi)}\frac{1}{ Q^2-M_\phi^2}
\langle\phi(Q,\varepsilon_\phi)|\rho(-k_1,\varepsilon)\pi(k_2)\rangle ,
\end{equation}
where $\varepsilon_\phi$ is the virtual $\phi$-meson polarization vector
and where a sum over all independent polarizations is implied. Noting that
the $\phi$ is nearly a pure $s\sbar$ state, so that $\bra{0}\ubar\gamma_\mu u
\ket{\phi}\approx 0\approx\bra{0}\dbar\gamma_\mu d\ket{\phi}$, one has (see the
Appendix
for details)
\begin{equation}
\bra{0}\,\sbar\gamma_\mu s\,\ket{\phi(Q, \varepsilon)}\approx -3\;
\bra{0}J_\mu^\sst{EM}\ket{\phi(Q, \varepsilon)} = -3\;\frac{M_\phi^2}{
f_\phi}\varepsilon_\mu ,
\end{equation}
where one obtains $f_\phi\approx 13 $ from an analysis of
$\Gamma(\phi\rightarrow e^+ e^-)$ $\cite{PDG}$. From the experimental value for
the
$\phi\to\rho\pi$ branching
ratio $\cite{PDG}$, one may obtain a value for the magnitude of the decay
amplitude,
$\langle\rho\pi|\phi\rangle$. From these inputs one obtains for the
transition form factor (see the Appendix)
\begin{equation}
\gropisff(Q^2)\Bigr\vert_{\phi-{\rm dom}} = \frac{\gropis}{ Q^2-M_\phi^2} ,
\end{equation}
with $|\gropis|\approx 0.20$ ${\rm GeV}^2$ $\cite{correct}$.

	In what follows, we consider additional contributions to
$\gropisff(Q^2)$ arising from loops, as in Fig. 3.

	The rationale for considering these contributions is similar to that
for including loops in studies of other form factors. From a dispersion
integral standpoint, $\gropisff (Q^2)$ receives contributions not only from
poles (identified with vector mesons), but also from the multi-meson,
intermediate-state continuum. For values of $Q^2$ sufficiently far from the
poles ({\it e.g.}, $M_\phi^2$), the continuum contributions need not be
negligible. In the present instance, the lightest allowed intermediate states
contain one pseudoscalar and one vector meson (parity  requires the
vanishing of the strong interaction $\pi\to $ two $K$ amplitude). Hence,
we will consider contributions involving one $K$ and one $K^\ast$ in the
intermediate state.

\section{ Effective Lagrangians}

In this section we derive  the effective interactions relevant
to the   calculation  of one-loop  contributions
to the matrix element  $\bra{\rho^a(k_1,\varepsilon)}\sbar\gamma_\mu
s\ket{\pi^b
(k_2)} $
and determine the effective coupling constants
associated with  them. In the derivation we use
an effective chiral Lagrangian, which takes into account   the global
symmetries
of QCD.
 This Lagrangian involves the octet of light pseudoscalar mesons
(the quasi-Goldstone bosons of spontaneously broken
chiral $SU_L(3)\times SU_R(3)$), the nonet of vector mesons, and external
gauge sources needed to obtain the relevant currents (electromagnetic and
strangeness) which play a role in our analysis.

The framework is that of the non-linear $\sigma$-model. The  quasi-Goldstone
fields
  are coordinates of the  coset space $SU_L(3)\times SU_R(3)/SU_V(3)\sim
SU(3)$,
and appear in the following form
\begin{equation}
U(x)=\exp\left(-i \frac{\Pi}{ F_0}\right),~~~~~~~\Pi=\sum_{a=1}^{8}\pi^a
\lambda^a ,
\end{equation}
where $\lambda^a$ are the Gell-Mann matrices normalized according to
$Tr(\lambda^a \lambda^b )=2\,\delta^{ab}$, $\pi^a$ are the members of the octet
of
pseudoscalar
quasi-Goldstone fields, and $F_0\sim 93$ MeV  is the pion
decay constant in the chiral limit. The  transformation properties
of the quasi-Goldstones are determined by the way  $U(x)$ responds to a
chiral transformation: $U(x)\rightarrow R U(x) L^{\dag}$, where L (R) belongs
to
$SU_L(3)$ ($SU_R(3)$).

The chiral transformation law  for the octet of vector mesons    is given by
\begin{equation}
V_\mu \rightarrow h\; V_\mu \;h^{\dag}
\end{equation}
where $V_\mu = V_\mu^a \lambda^a $, $u(x)\equiv\sqrt{U(x)}$, and
$h=h(L,R,u(x))$
is determined by the equations
\begin{equation}
L\;u(x) = u^\prime (x) \, h ~~~~~~~~~~~~~R\;u^{\dag} (x) = u^{\prime^{\dag}}
(x)
\, h
\end{equation}
where $u^\prime$ results from the action of the chiral transformation on $u$.

In order to build a chirally invariant Lagrangian for the octet
of vector mesons we further need to introduce a covariant derivative
\begin{equation}
\nabla_\mu = \partial_\mu-i\,\Gamma_\mu ~~~~~~~~~~~~
\Gamma_\mu = \frac{i}{ 2}\,(u^{\dag} \partial_\mu u+u\partial_\mu u^{\dag} ),
\end{equation}
which operates as follows on $V_\mu$: $\nabla_\mu V_\nu=\partial_\mu V_\nu
-i\left[ \Gamma_\mu, V_\nu \right]$. Similarly, we require
the axial vector connection
\begin{equation}
\omega_\mu=\frac{i}{ 2}\,(u^{\dag}\partial_\mu u-u\partial_\mu u^{\dag}).
\end{equation}
Both $\nabla_\mu$ and  $\omega_\mu$  have the same chiral transformation law as
$V_\mu$.

In addition, we will consider two external source  gauge fields,
the electromagnetic potential $A_\mu$ and  the potential ${\cal S}_\mu$
which couples to the strangeness vector current. In this way one can
obtain the most general
form of the respective currents in the effective theory. At the appropriate
stage we will show how they enter in the effective Lagrangian.

The piece of the effective Lagrangian containing the kinetic and mass terms
of the vector meson octet is given by
\begin{equation}
{\cal L}_V=-\frac{1}{8}\;{\rm Tr}(V_{\mu\nu}V^{\mu\nu} )+ \frac{M_V^2}{4}\;{\rm
Tr}(V_\mu V^\mu )\ \ \ ,
\end{equation}
where
$V_{\mu\nu}=\nabla_\mu V_\nu-\nabla_\nu V_\mu$.
The Lagrangian ${\cal L}_V$  only contains terms with two vector mesons ($V$)
and even numbers
of quasi-Goldstone bosons (P). There is no SU(3) breaking at this
 level in the interactions. For simplicity, we will only include the dominant
SU(3) breaking effects which are generated by the mass splittings  in the
octets of pseudoscalar and vector mesons.

The vertex of type $VVP$ needed in our analysis is specifically
$\rho K^\ast K $. In order to determine the corresponding coupling constant
we need to consider  in addition the $VV^0 P$  vertices, where ${\rm V}^0$
is the singlet component of the nonet of vector mesons.
The corresponding effective Lagrangians  at leading order in chiral power
counting are of ${\cal O}(p)$, where $p$ denotes the generic small momentum
carried by the quasi-Goldstone
bosons, and read
\begin{eqnarray}
 {\cal L}_{VVP}&=& R_8 \; \epsilon_{\mu\nu\rho\sigma} \;{\rm Tr}(\{\nabla^\mu
\,V^\nu ,\, V^\rho \}\omega^\sigma ) \nonumber \\
{\cal L}_{VV^0 P}&=& R_0 \;\epsilon_{\mu\nu\rho\sigma}
\;{\rm Tr}(\nabla^\mu \,V^\nu , \omega^\rho )V^{0 \sigma}\ \ \ .
\end{eqnarray}
Here,  we have used the identity $ \epsilon_{\mu\nu\rho\sigma} \nabla^\rho
\omega^\sigma =0$
to simplify the expressions.  For the vertices of interest we need only keep
the
first term in the expansion of
$\omega^\mu$: $\omega^\mu = \frac{1}{ 2\,F_0} \partial^\mu \Pi +{\cal
O}(\Pi^3)$.
The only observable strong interaction process of this
type is $\phi\rightarrow\rho\pi$. In order to be able to determine
$R_8$, we must simultaneously pin down $R_0$, and this requires
further information, which we obtain by considering the radiative decays
$V\rightarrow P\gamma$ supplemented with the  hypothesis of vector meson
dominance. From this   analysis, presented in the Appendix,
we obtain  $R_8=0.22\pm0.08$.

The vertices of type $VPP$ we need are of the type $K^\ast K \pi$. They are
determined in terms of a single effective coupling, and the corresponding
effective Lagrangian is
\begin{equation}
{\cal L}_{VPP}= \xi_8 \;{\rm Tr}(\nabla_\mu V_\nu \;[\omega^\mu ,\omega^\nu ] )
\ \ \ ,
\end{equation}
where $\xi_8$ can be determined from either of the following two decay widths:
\begin{eqnarray}
  \Gamma(\rho^0 \rightarrow \pi^+ \pi^- )&=& \frac{\xi_8^2 }{ 192\pi F_{0}^4
}\,M_\rho^2 \,(M_\rho^2 -4 M_\pi^2 )^{3/2} \nonumber \\
\Gamma(K^{\ast \,+} \rightarrow K^+ \pi^0 )&=&
\frac{\xi_8^2 }{ 96\pi F_0^4 }\,E_K\,E_\pi \,k_f , ~~~~~~
k_f= \frac{1}{ 2M_{K^\ast }}\sqrt{\lambda[M_{K^\ast }^2, M_K^2 ,M_\pi^2 ]}
\ \ \ ,
\end{eqnarray}
where $\lambda[x,y,z]=x^2+y^2+z^2-2 x y- 2 x z- 2 y z$. The first width gives
$\xi_8=0.175$ and the second gives  $\xi_8=0.140$. Notice
that we
have used the same decay constant for $\pi$ and $K$ mesons. Under the
assumption
of
 $SU(3)$  symmetry in the interactions,
the use of  either value of $\xi_8$ is justified.

Now we turn to those terms in the effective Lagrangian which determine the
electromagnetic and strangeness
currents. The electric charge operator  contains only octet pieces and is
explicitly given by
$\hat{Q}= \frac{e}{ 2}(\lambda^3 + \frac{1}{\sqrt{3}} \,\lambda^8
)=\frac{e}{3}\,{\rm diag}(2,-1,-1)$, while the strangeness charge operator
contains a singlet and an octet piece and reads $\hat{S}= \frac{1}{
3}(1-\sqrt{3}\;\lambda^8 )={\rm diag}(0,0,1)$.
In the following, $\hat{q}$ denotes either of these charge operators, and
$v_\mu$ represents  either of the two
source fields $A_\mu$ and ${\cal S}_\mu$.

Let us first consider the $V\rightarrow v_\mu$  amplitudes relevant for the VMD
analysis.
The effective Lagrangian contains an octet and a singlet piece
\begin{equation}
{\cal L}_{Vv}=C_8\,v^\mu \;{\rm Tr}(\hat{q} V_\mu)+C_0\,v^\mu \;{\rm
Tr}(\hat{q})\,V^0_\mu\ \ \ .
\end{equation}
The leptonic widths of $\rho^0$, $\omega$ and $\phi$ determine $C_8$ according
to:
\begin{equation}
\Gamma(V\rightarrow e^+ e^-)=\frac{4 \pi}{ 3} \alpha^2 \; \frac{M_V}{f_V^2}
\ \ \ ,
\end{equation}
where
\begin{eqnarray}
f_\rho&= &\frac{\sqrt{2} \,M_\rho^2 }{ C_8}\nonumber\\
 f_\omega &=& \frac{\sqrt{6}\,M_\omega^2}{
{\cos \theta} \,C_8}\nonumber\\
f_\phi &=& \frac{\sqrt{6}\,M_\phi^2}{\sin \theta \,C_8}\ \ \ .
\end{eqnarray}
In practice, the $\omega - \phi$ mixing angle is taken to be that of ideal
mixing,
${\tan \theta}=\sqrt{2}$,
which leads to $C_8\simeq 0.16$ $\rm{GeV}^2$.  In order to determine the
transition
mediated by the strangeness current, we also need  to know the singlet coupling
$C_0$.  As there is no  direct experimental
determination of this coupling,   one must rely on some hypothesis. By assuming
exact OZI suppression in the $\omega$-to-vacuum current matrix elements
({\it i.e.}, $\bra{0}\sbar\gamma_\mu s\ket{\omega}=0 $), we obtain
$C_0=- C_8/\sqrt{3}=-0.092$ $\rm{GeV}^2$.

The  transitions  within the    octet of vector mesons mediated by the currents
are obtained
 by minimal substitution
into the chiral covariant derivative $\nabla_\mu$ in Eq. (12), plus a gauge
invariant term
involving the field strength tensor of the gauge field. Both electromagnetic
and
strangeness current
transitions are  determined by the  following  effective Lagrangian:
\begin{equation}
{\cal L}_{VVv}= \frac{i}{ 4}\,{\rm Tr}(V_{\mu\nu}\,[\hat{q},V^\nu ]) \; v^\mu
+i \, \frac{ z}{2}\;{\rm Tr}([V_\mu,V_\nu ]\;\hat{q}) \;v^{\mu\nu}\ \ \ .
\end{equation}
Only one  new  unknown effective   coupling  ($ z$)   enters, which is related
to the magnetic moments of the vector mesons.
As we mention in the Section V, the lack of experimental access to $z$ does not
affect our results, as it appears only in a loop diagram which turns out to be
subdominant (diagram 3c).

Vertices of type $VPPv$ are obtained from Eq. (14) by minimal substitution in
$\nabla_\mu$
and $\omega_\mu$, and by a term proportional to the field strength tensor of
the
gauge field.
The effective Lagrangian reads (for the sake of simplicity we keep only terms
relevant to our calculation):
\begin{eqnarray}
  {\cal L}_{VPPv}&=&-i \, \frac{\xi_8}{ 4\,F_0^2 }\;{\rm Tr}\left(\partial^\mu
V^\nu ([\partial_\mu\Pi,\;
[\hat{q},\Pi ]\,]v_\nu-\mu \leftrightarrow\nu)\right.\nonumber \\
&+&\left.v^\mu [\hat{q},\,V^\nu ]\,[\partial_\mu \Pi,\,\partial_\nu
\Pi]\right)\nonumber \\
&+&i\, \frac{\zeta_8}{ 4\,F_0^2 } \;{\rm Tr}\left(\partial^\mu \,V^\nu \,
[[\hat{q},\Pi],\Pi]\right)\;v_{\mu\nu}+...\ \ \ .
\end{eqnarray}
The new coupling constant $\zeta_8$ is fixed by considering the radiative
decay
$\rho^0 \rightarrow\pi^+ \pi^- \gamma$.
Using the expression for the partial width
\begin{eqnarray}
\Gamma(\rho^0 \rightarrow \pi^+ \pi^- \gamma)&= &\frac{\alpha\,M_\rho}{
24\,\pi^2 \,F_0^4}
\int_{M_\pi}^{M_\rho}\,dE_1\,\int_{M_\pi}^{M_\rho-E_1}\,dE_2\,\nonumber \\
&\times& \Theta\left[4(E_1^2 -M_\pi^2)(E_2^2  -M_\pi^2 )  \right. \nonumber\\
&-& \left.(M_\rho^2 +2M_\pi^2 -2(E_1+E_2)M_\rho +2\,E_1\,E_2)^2\right]
\nonumber
\\
&\times& \left(2 \zeta_8\,M_\rho-(2\zeta_8-\xi_8)\,(E_1 +E_2)\right)^2
\end{eqnarray}
and using  the  previously determined value $\xi_8 =0.175$,  we obtain two
solutions: $\zeta_8 =0.156$ and $\zeta_8 =-0.082$.
Unfortunately, there are no further measured observables to discriminate
between
these two solutions. In our results we will include both
possibilities.

The final  vertices we need are those of the type  $VVPv$. As in the previous
case, we only need those involving
members of the meson octets. They are entirely determined by minimal
substitution into
eq. (13), and the relevant pieces read:
\begin{eqnarray}
 {\cal L}_{VVPv} &=&-i\, \frac{R_8 }{ 2\,F_0} \;\epsilon_{\mu\nu\rho\sigma} \;
{\rm Tr}\left(v^\mu \{[\hat{q},\, V^\nu ],\, V^\rho \}\partial^\sigma
\Pi\right.
 \nonumber \\
&+&\left. v^\sigma \{\partial^\mu V^\nu ,\, V^\rho\} [\hat{q},\,\Pi]\right)
\ \ \ .
\end{eqnarray}

{}From the effective Lagrangians given in this section it is straightforward to
derive the
Feynman rules and calculate the one-loop diagrams in the following section.

\section{  One Loop Calculation}

In this section we discuss some  salient features of the calculation
of the one-loop
contributions to $\bra{\rho^0(k_1,\varepsilon)}\sbar\gamma_\mu
s\ket{\pi^0(k_2)}$, which on grounds of isospin symmetry
are the same as the contributions    to
$\bra{\rho^+(k_1,\varepsilon)}\sbar\gamma_\mu s\ket{\pi^+(k_2)}$.

There are only four diagrams, depicted in Fig. 3,  which contribute at one-loop
order.
As expected, all diagrams are ultraviolet divergent. Diagrams 3a, b  and d are
quadratically divergent, while diagram
3c is only logarithmically divergent.  At this point we must warn that, quite
in general,
 the present calculation
shows lack of an appropriate chiral expansion  as an expansion in the number of
loops; this is due to the presence of
a heavy meson ($K^{\ast}$) in the loop.  Hence, in contrast to calculations
involving light mesons or stable heavy baryons carried out using chiral
perturbation theory, in the present case multi-loop and one-loop
contributions might have comparable magnitudes.

 We regulate the loops by introducing a hadronic form factor
at the vertices. We employ a form factor, rather than a momentum
cut-off as is conventional in chiral loop calculations, for two reasons.
First, in the conventional framework, the cut-off dependence of a physical
amplitude is removed by introducing additional, higher-dimension operators
into the effective Lagrangian whose coefficients are determined from
measurements of one or more observables (often using SU(3) or other
symmetry relations). In the present case, however, we are concerned with
a matrix element  of an operator which contains both SU(3) octet and
singlet components: $\sbar\gamma_\mu s=V_\mu^{(0)}+(2/\sqrt{3})V_\mu^{(8)}$.
Symmetry arguments only allow one to determine coefficients of
higher-dimension octet operators from, {\it e.g.}, measured hyperon
semi-leptonic decay rates and low-lying baryon vector current form factors.
There exists insufficient experimental information to fix the coefficients
of higher-dimension singlet operators. Consequently, the predictive
power of claculations employing a conventional momentum cut-off
procedure is limited in the case of strangeness matrix elements. The
use of an alternative regulator, such as a hadronic form factor,
allows one to overcome this limitation, although at the cost of
introducing additional model-dependence into the calculation.

A second reason for the use of a form factor is that it affords one
a natural means of preserving the vector current Ward identities
(see below), which are violated by the use of a simple momentum
cut-off procedure.

For purposes of simplicity, we choose the form factor to depend only on
$k^2$, the momentum
squared of the  virtual K-meson, and use the same cut-off
mass parameter for all types of vertices.
We expect this choice to be of little significance concerning the generality of
our results.
The hadronic form factor is taken to be
\begin{equation}
F(k^2)=\,\frac{M_K^2-\Lambda^2 }{ k^2-\Lambda^2+i\epsilon}\ \ \ ,
\end{equation}
where the scale $\Lambda$ will be chosen within a reasonable range as
discussed below. For $k^2=M_K^2$, one has $F(M_K^2)=1$. Hence, this
choice for the form factor is consistent with the values of the
mesonic coupling constants $(R_0, R_8, \xi_8)$ extracted from
on-shell amplitudes.

\def\mk{{M_K}}
\def\mks{{M_K^2}}
\def\msk{{M_{K^\ast}}}
\def\msks{{M_{K^\ast}^2}}

The implementation of this form factor in the diagrams where the current
is inserted in the meson line (3c,d)
is performed by the following  straightforward
procedure: if $G(\mks, \msks)$ denotes the diagram for point-like
hadronic vertices ($F(k^2)\equiv 1$), then the corresponding diagram
with $F(k^2)$ as given in Eq. (23) is
\begin{equation}
\hat{\cal F}(\Lambda)\left[G(\mks, \msks)-G(\Lambda^2, \msks)\right],
\end{equation}
where the operator $\hat{\cal F}$ is given by
\begin{equation}
\hat{\cal F} (\Lambda) = - (\Lambda^2-M_K^2)^2\;\frac{\partial }{ \partial
\Lambda^2} \frac{1}{ M_K^2-\Lambda^2}\ \ \ .
\end{equation}

We note that, on general grounds, the introduction of
an additional momentum-dependence at the vertices also requires the
inclusion of new \lq\lq seagull" vertices in order to maintain
gauge invariance $\cite{gross,ohta,mike4}$. In the present case,
these seagull terms generate additional
terms in the $VPPv$ and $VVPv$ interactions given in Eqs. (20) and
(22). Although there exists no unique prescription for maintaining
gauge-invariance in the presence of hadronic form factors, we follow
the minimal prescription of Refs.~$\cite{gross,ohta,mike4}$ to derive our
seagull
terms. To this end, one may consider the form factors appearing at the
hadronic vertices as arising from a co-ordinate space interaction of
the form
\begin{equation}
\Psi(\phi_1,\ldots,\phi_k)_\lambda F(-\partial^2)\partial^\lambda\Pi
\ \ \ ,
\end{equation}
where $\Pi$ is the octet of pseudoscalar meson fields defined in Eq.~(7),
$\Psi_\lambda$ is a Lorentz vector constructed from the other pseudoscalar
and vector meson fields, and $\partial^2$ is the D'Alembertian. Transforming
to momentum space yields the same hadronic vertices as generated by the
effective Lagrangians in Eqs.~(13-14) multiplied by the form factor
$F(k^2)$, where $k$ is the momentum of the pseudoscalar meson associated
with the field $\Pi$. The gauge invariance of this interaction can be
restored by making the minimal substitution $\partial_\mu\to D_\mu=
\partial_\mu-i\hat q v_\mu$, where $\hat q$ is either the EM or strangeness
charge operator and $v_\mu$ is the corresponding source field. Expanding
the resultant interaction to first order in $v_\mu$ yields the seagull
interaction
\begin{equation}
\Psi(\phi_1,\ldots,\phi_k)_\lambda\Bigl\{ v\cdot(Q+2i\partial)
\left[{F(-\Delta^2)-F(-\partial^2)\over(-\Delta^2)-(-\partial^2)}\right]
\partial^\lambda\left[\hat q, \Pi\right]
-iv^\lambda F(-\Delta^2)\left[\hat q, \Pi\right]\Bigr\}\ \ \ .
\end{equation}
Here, $\Delta^2=\partial^2-2iQ\cdot\partial-Q^2$, where $Q$ is the
momentum carried by the source. Transforming to momentum space,
replacing the source by the associated vector boson polarization vector
$\varepsilon_\mu(Q)$, and taking the specific form for the form factor
given in Eq.~(23) leads to the vertex structure
\begin{equation}
-i\tilde\Psi_\lambda\left\{\mp F(k^2)\left[{(Q\pm 2k)^\mu\over
(Q\pm k)^2-\Lambda^2}\right]k^\lambda
\varepsilon_\mu(q)
+F((Q\pm k)^2)\varepsilon^\lambda(q)\right\}\left[\hat q, \lambda^a\right]\
\ \ \ ,
\end{equation}
where the  SU(3) index \lq\lq $a$" is associated with the pseudoscalar
meson carrying the momentum $k$, $\tilde\Psi_\lambda$ is the
momentum-space form for the $\Psi_\lambda$ (without the field operators),
and the upper (lower) sign corresponds to an incoming (outgoing) pseudoscalar
meson.

Inclusion of this \lq\lq minimal" seagull vertex (via diagrams 3a,b) is
sufficient to preserve the gauge invariance of the loop calculation in the
presence of hadronic form factors (for a demonstration of this feature
for the case of meson-baryon loops, see
Refs.~$\cite{mike4,forkel,koepf}$). One may,
of course, include additional transverse seagull terms, which are
separately conserved and which, therefore, do not modify the gauge-invariance
of the calculation. This possibility introduces a degree of model-dependence
into the form factor calculation. However, for purposes of the present work,
where we seek to set the scale of, and estimate the theoretical uncertainty
associated with, loop contributions, the use of the minimal prescription is
sufficient.

In general, the inclusion of hadronic form factors also destroys the chiral
invariance of the calculation, since the interaction in Eq.~(26) is not
invariant under a local chiral rotation. Restoration of chiral invariance
would necessitate replacement of the derivatives $\partial_\mu$ acting
on $\Pi$ by the chiral covariant derivative $\nabla_\mu$ introduced in
Eq.~(10). Since the vector connection $\Gamma_\mu$ appearing in $\nabla_\mu$
can be expanded in a power series in the pseudoscalar meson field $\Pi$,
this prescription for maintaining chiral symmetry would generate new
chiral seagull vertices containing additional pseudoscalar mesons.
The chiral seagull vertex of lowest order in $1/F_0$
would contain two additional pseudoscalar meson fields and, consequently,
would first contribute to the transition form factor at two-loop order.
Since we restrict our attention to one-loop results, we do not consider
contributions from these chiral seagull interactions.

 In the notation introduced in the previous section, the VMD
piece of this form factor is given by
\begin{equation}
\frac{1}{M_\rho}\,g_{\rho \pi}^{(s)}(Q^2=0)\Bigr\vert_{\phi-{\rm dom}}=
- \frac{3}{F_0} \;\left(\frac{4}{\sqrt{3}} R_8\,\sin\theta-
R_0\,\cos\theta\right)
 \frac{1}{ f_\phi}.
\end{equation}
It is convenient to use the  measured rate for $\phi \rightarrow \rho \pi  $,
which gives:
\begin{equation}
\frac{1}{M_\rho}\,g_{\rho \pi}^{(s)}(Q^2=0)\Bigr\vert_{\phi-{\rm dom}}=
 - \frac{3}{f_\phi}\; G_{\rho \phi\pi}^{{\rm Phen}},
\end{equation}
with  $G_{\rho \phi\pi}^{{\rm Phen}}=1.08\;\, {\rm GeV ^{-1}}$. As shown in the
Appendix, we can use Eq.~(30)  to
determine the combination of $R_0$ and $R_8$ which appears in Eq.~(29).

The one-loop contributions are now denoted by
 $g_{\rho \pi}^{(s)}(Q^2, k_1^2, k_1\cdot Q)\Bigr\vert_{1-loop}^{(j)}$,
where $j=a,b,c,d$ refers to
the contributions from the different diagrams, and where the
 dependence on $k_1^2$ and $k_1\cdot Q$ allow for off-shell initial and
final state mesons. In order to give the expressions
of these
contributions in  a convenient form, we introduce an  operator to project out
the coefficients accompaning terms proportional to $k_1^\alpha$ or $Q^\alpha$
resulting from the
loop integrals:
\begin{equation}
\hat{P}_{\alpha}(p,q)=\frac{p^2\,q_\alpha-p\cdot q\,p_\alpha}{p^2\,q^2-(p\cdot
q)^2}.
\end{equation}
We then have that $\hat{P}_{\alpha}(p,q) p^{\alpha}=0$,
$\hat{P}_{\alpha}(p,q) q^{\alpha}=1$.

The expressions  for the different
diagrams then
read as follows:
\begin{eqnarray}
\frac{1}{M_\rho}\,g_{\rho \pi}^{(s)}(Q^2,k_1^2,k_1 .
Q)\Bigr\vert_{1-loop}^{(a)}&=&
\frac{4\,R_8}{F_0^3}\;\int
\frac{d^4k}{(2\pi)^4}\;\frac{1}{(k^2-M_K^2)((k_1-k)^2-M_{K^\ast}^2)}\nonumber\\
&\times & \left\{\frac{1}{2}(\xi_8-\zeta_8)\,F(k)^2\,k^2
- \xi_8F(k)F(k+Q)\right.\nonumber\\
&\times& \left.\left[\frac{k^2}{4}\,(-1+2\frac{k\cdot
(k_1-k)}{k^2-\Lambda^2})
+ \hat{P}_\alpha(k_1,Q)k^\alpha
\;(k_1+Q)\cdot(k_1-k)\right]\right\} \nonumber\\
\frac{1}{M_\rho}\,g_{\rho \pi}^{(s)}( k_1^2, k_1\cdot
Q)\Bigr\vert_{1-loop}^{(b)}&=&
\frac{2\,R_8}{F_0^3}\;\xi_8\;\int
\frac{d^4k}{(2\pi)^4}\;\frac{F(k)F(k-Q)}{(k^2-M_K^2)((k_1+Q-k)^2-M_{K^\ast}^2)}
\nonumber\\
&\times & \left\{k\cdot
(k_1+Q-k)\;(1-\frac{1}{2}\frac{k^2}{k^2-\Lambda^2})\right.\nonumber\\
&-& \hat{P}_\alpha(k_1,Q) k^\alpha \; \left( (k_1+Q)^2+k \cdot
(k-2(k_1+Q))\right)\nonumber\\
&-& \left. \hat{P}_\alpha(Q, k_1) k^\alpha \; \left(2 (k_1+Q)^2+k \cdot
(k-3(k_1+Q))\right)\right\}\nonumber\\
\frac{1}{M_\rho}\,g_{\rho \pi}^{(s)}(Q^2, k_1^2, k_1\cdot
Q)\Bigr\vert_{1-loop}^{(c)}&=&
-\frac{4 \, R_8 \,\xi_8}{F_0^3}\;\int
\frac{d^4k}{(2\pi)^4}\;F^2(k)\nonumber\\&\times&
\frac{1}{(k^2-M_K^2)((k_1-k)^2-M_{K^\ast}^2)((k_1+Q-
k)^2-M_{K^\ast}^2)}\nonumber\\
&\times & \left\{(1+4 z)\,\hat{P}_\alpha(Q,k_1)k^\alpha\;
\left[Q . k \;(p_1+Q)-Q\cdot(k_1+Q)\;k\right]\cdot (k_1-k)\right.\nonumber\\
&+&\left. \frac{k^2}{4}\;\left[ -(1+4 z)\;(k_1+Q)^2+(3+4
z)\;k\cdot(k_1+Q)\right]\right\}\nonumber\\
\frac{1}{M_\rho}\,g_{\rho \pi}^{(s)}(Q^2, k_1^2, k_1 .
Q)\Bigr\vert_{1-loop}^{(d)}&=&
\frac{4 \, R_8\,\xi_8}{F_0^3} \;\int \frac{d^4k}{(2\pi)^4}\;
\frac{F^2(k)\;
k^2\;(k_1-k)^2}{(k^2-M_K^2)((k+Q)^2-M_{K})((k_1-k)^2-M_{K^\ast}^2)}
\end{eqnarray}

For purposes of comparison we also   consider the one-loop contributions to
the  matrix elements of the electromagnetic current
 $\bra{\rho(k_1,\varepsilon)}J^\sst{EM}_\mu  \ket{\pi(k_2)}$ and
$\bra{K^{\ast}(k_1,\varepsilon)}J^\sst{EM}_\mu  \ket{K(k_2)}$.
The one-loop contributions in the case of the electromagnetic current are
related to those already
obtained above
in a  simple manner. With obvious notation, ($f$ and $d$ are structure
constants
of SU(3)),
\begin{eqnarray}
g_{\rho \pi}^{(\gamma)}\Bigr\vert_{1-loop}&=&-\frac{e}{2} g_{\rho
\pi}^{(s)}\Bigr\vert_{1-loop}\nonumber \\
\frac{1}{M_{K^\ast}}\,g_{K^{\ast}K}^{(\gamma)}\Bigr\vert_{1-loop}&=&-\frac{e}{M_
\rho}\,\left(\sum_{i=1}^{3}d_{K^\ast i b} f_{K i c} (f_{3 b c}+
\frac{1}{\sqrt{3}} f_{8 b c})  g_{\rho
\pi}^{(s)}\Bigr\vert_{1-loop}(M_{K^\ast}\rightarrow M_\rho)\right.\nonumber\\
&+& \sum_{i=4}^{7}d_{K^\ast i b} f_{K i c} (f_{3 b c}+
\frac{1}{\sqrt{3}} f_{8 b c})  g_{\rho
\pi}^{(s)}\Bigr\vert_{1-loop}(M_{K}\rightarrow M_\pi)\nonumber\\
&+&\left.  d_{K^\ast 8 b} f_{K 8 c} (f_{3 b c}+
\frac{1}{\sqrt{3}} f_{8 b c})  g_{\rho
\pi}^{(s)}\Bigr\vert_{1-loop}(M_{K^\ast}\rightarrow M_\omega) \right),
\end{eqnarray}
where the cases of charged and neutral kaons must be considered separately. It
is noteworthy that  the   one-loop
contributions to $g_{\rho \pi}^{(\gamma)}$ are   due purely  to  strange
particles in the loop.

\section{ Results}

\def\rgamvp{{R_{VP}^{(\gamma)}}}
\def\rgamropi{{R_{\rho\pi}^{(\gamma)}}}
\def\rgamkk{{R_{K^\ast K}^{(\gamma)}}}
\def\rsropi{{\tilde R_{\rho\pi}^{(s)}}}

In this section we present and discuss the results for the one-loop
contributions to the transition  matrix elements of both  electromagnetic and
strangeness currents. Throughout we use  $R_8=0.27$,  $\xi_8=0.18$,  $z=0$,
and,
$\zeta_8=0.16$ (case 1) and  $\zeta_8=-0.08$  (case 2).
Since diagram (c) turns out to give a  modest contribution, the choice
$z=0$ has little impact on the  results.  If we instead use $\xi_8=0.14$,
as obtained by
considering $K^\ast$ decay, the results obtained for $\zeta_8$
and the final results of the loop calculation  are only   slightly affected.

We discuss first the case of the electromagnetic current matrix elements.
To this end, it is convenient to define the magnitude of the ratio
of the one-loop contribution to the form factor to the measured form
factor (\lq\lq Phen")
\begin{equation}
\rgamvp\equiv\left|{g_{VP}^{(\gamma)}({\rm loop})\over
g_{VP}^{(\gamma)}({\rm Phen})}\right| \ \ \ .
\end{equation}
The sign of the ratio is  undetermined,
since the sign of the low energy coupling constants involved cannot be
established
from available observables. Hence, we consider only the absolute value.
The ratio $\rgamvp$ is shown in Figs.
4 and 5 for the cases of $\rho$-to-$\pi$ and $K^\ast$-to-$K$  radiative
transitions respectively.

In each case, the choice of a \lq\lq reasonable
range" for the mass parameter $\Lambda$ was dictated by two criteria.
First, in order to maintain consistency with our use of VMD in extracting
some of the coupling constants from radiative transitions, we require
$\Lambda$ to fall within a range such that $R^{(\gamma)}_{\rho \pi}\leq 1/2$
($R^{(\gamma)}_{\rho \pi}=1/2$
corresponds to loop and vector meson poles given equal contributions
if the signs of the contributions are the same).
This condition
gives an upper bound of roughly one GeV. Second, to obtain a lower bound
on $\Lambda$, we refer to a \lq\lq cloudy bag" picture of hadrons in
which the pseudoscalar Goldstone bosons live outside a hadronic bag
containing quarks. In this picture, the virtual meson must have a
wavelength longer than the bag radius, so as to be unable to penetrate
the bag interior. From this requirement, we obtain a lower bound of
$\Lambda\sim 1/R_{\rm bag}\approx 0.2\ \hbox{GeV}$ for a bag radius of
one Fermi. For this choice of $\Lambda$, the form factor in Eq. (23)
will suppress contributions from virtual pseudoscalar mesons with
wavelengths shorter than one Fermi. We emphasize that although we
do not perform this calculation
within the cloudy bag framework, we simply turn to that picture to obtain
a physical argument for a reasonable lower bound on the cutoff mass.

For $\Lambda$ falling within our \lq\lq reasonable range", the diagrams
contributing to the $\rho\pi\gamma$ form factor
display a zero at $\Lambda = \mk$. This zero results from the
numerator in Eq. (23), which was chosen to give the normalization
$F(\mks)=1$. Hence, this
zero should be taken as an un-physical artifact of the choice of form
factor. We believe that the values of $\rgamvp$ at $\Lambda\sim 0.2\
\hbox{\rm GeV}$ give a realistic lower bound, since this value on
$\Lambda$ is sufficiently far from the artificial zero at $\mk$. In
order to check this assumption, we also computed the diagrams with
a slightly different form factor, replacing the numerator of Eq. (23)
with $-\Lambda^2$. Such a form factor  also displays pointlike behavior
($F(k^2)=1$) as $\Lambda\to\infty$.
The resulting values of $\rgamvp$
for $\Lambda$ in the vicinity of $\mk$ do not differ significantly
from those obtained with the form in Eq. (23) and $\Lambda\sim 0.2\
\hbox{\rm GeV}$ (see Fig. 6).  In the case of the $K^\ast\to K\gamma$
form factor, no zero
appears at $\Lambda=\mk$ since loops involving virtual pions
also contribute.
The latter enter with hadronic form factors normalized to
unity at $k^2=\mpis$
and, thus, do not vanish at $\Lambda=\mk$.

 The one-loop contribution in  the  $\rho\pi\gamma$  case is about
5 \% when  $\Lambda\sim 0.7$ GeV, and grows to 25-50\% for $\Lambda\sim 1$
GeV. Thus, we take 1 GeV as an upper bound for our reasonable range for
$\Lambda$. For the $K^{\ast}K\gamma$ form factor, the loops containing
a $\pi$ and a $K^{\ast}$ in the intermediate state are the dominant ones.
The loop contributions are relatively more important
than those which enter the
$\rho\pi\gamma$ form factor, ranging from 50-250\% of the experimental
value as $\Lambda$ varies from $0.2\to 1$ GeV.
The relatively large one-loop contributions are an  indication that the
assumed validity of VMD for radiative transitions
is most likely affected by substantial  corrections.
Notice that our analysis relies
on VMD to determine some low energy constants, like $R_8$ and $R_0$
(see the Appendix). The large loop contribution to the radiative
$K^{\ast}$ decays suggests that one ought to perform a self-consistent
fitting procedure, in which both VMD and one-loop amplitudes are
included in the determination of $R_8$ and $R_0$. Such an analysis
goes beyond the scope of our present objective, however. We seek
only to determine the magnitude of, and quantify the degree of
theoretical uncertainty associated with, loop contributions to
$g_{\rho\pi}^{(s)}$ rather than to arrive at a definitive numerical
prediction. From this standpoint, we do not expect that a carrying
out a self-consistent fitting procedure would significantly modify our
conclusions.

For the strangeness vector current transition form factor, we must define
a somewhat different ratio since $g_{\rho \pi}^{(s)}(Q^2)$ has not been
measured. We compute instead ratio $\rsropi$ of one-loop to
vector meson dominance contributions
\begin{equation}
\rsropi\equiv\left|{g_{\rho \pi}^{(s)}({\rm loop})\over
g_{\rho \pi}^{(s)}({\rm VMD})}
\right|\ \ \ .
\end{equation}
The contributions of the different diagrams to this ratio are all of the same
sign. The absence of possible cancellations between diagrams renders our
results
less sensitive to the precise numerical values of the low energy constatnts or
choice of hadronic form factor than would be the case in the presence of
cancellations.  A similar situation occurs in the ratio
$R^{(\gamma)}_{\rho \pi}$. In the case of the ratios $\rgamkk$
there are contributions of different signs. However, the loops containing pions
are dominant and  all of the same sign.

The results for the ratio $\rsropi$, shown in Fig. 6, turn out to
be   large even for very modest values of the cutoff mass. The reason
is that $g_{\rho \pi}^{(s)}(Q^2=0)\Bigr\vert_{\phi-{\rm dom}}$
is very small
( a factor three smaller than $1/e\;g_{\rho\pi}^{(\gamma)\,{\rm Phen}}$ ),
while the one-loop
correction is a factor two larger than  $1/e\;g_{\rho\,\pi
}^{(\gamma)}\Bigr\vert_{1-loop}$; thus,
the corresponding ratio is almost a factor {\it six} larger than in
$\rho\pi\gamma$ case. Consequently, for $\Lambda\sim 1$ GeV, the strangeness
$\rho-\pi$
transition matrix element can be as large in magnitude as the corresponding
EM transition matrix element, assuming the loop and $\phi$-pole contributions
enter with the same sign (uncertain at present). For small values of the
form factor mass ($\Lambda\sim 0.2$ GeV), however, the loop correction
to the $\phi$-pole contribution is 50\% at most, in which case
$g_{\rho\pi}^{(s)}(0)$ is no more than half as large as
$g_{\rho\pi}^{(\gamma)}(0)$.

In addition to giving the $\Lambda$-dependence of $\rsropi$, the
curves in Fig. 6 also illustrate the sensitivity of our results to
other parameters which enter.
As in the case of $g_{\rho\pi}^{(\gamma)}(0)$, the dependence of the
strangeness
form factor on the choice of low-energy constant $\zeta_8$ is modest:
the lower set of curves (case 1) and upper set (case 2) differ by less
than a factor of two over our reasonable range for $\Lambda$. Similarly,
the dependence on the $\rho$ and $\pi$ virtuality is negligible, as
a comparison of the solid curves ($k_1^2=\mros$, $k_2^2=\mpis$) and
dashed curves ($k_1^2=0$, $k_2^2=0$) indicates. For this reason,
we conclude that a nuclear MEC calculation carried out using
$k_1^2=\mros$ and $k_2^2=\mpis$ in $g_{\rho\pi}^{(s)}$ rather than
allowing these momenta to vary introduces negligible error.

The dotted curve in Fig. 6 also indicates the sensitivity to hadronic
form factor normalization. This curve gives the result when one uses
$F(k^2)=-\Lambda^2/(k^2-\Lambda^2)$ rather than the form in Eq.~(23). The
results do not differ significantly except in the vicinity of $\Lambda=
\mk$ as expected (see above discussion). In this region, the values of
the dotted curve are close to those of the solid curves at $\Lambda=
0.2$ GeV, confirming our assertion that the latter give a reasonable
lower bound on $\rsropi$.

Finally, we refer to the dash-dotted curve, which gives the ratio
$\rsropi$ (for case 1) at the kinematics of the
approved CEBAF experiment $\cite{cebaf1}$, $|Q^2|=0.6\ (\hbox{GeV}/c)^2$.
In this
case the $\phi$-pole contribution to $g_{\rho\pi}^{(s)}(Q^2)$ is down from its
value at the photon point by $(1-Q^2/m^2_\phi)^{-1}\approx 0.6$. The ratio
$\tilde R_{\rho\pi}^{(s)}$, on the other hand, is essentially unchanged
from its value at the photon point. Thus, we would expect the loop
contributions to modify the strangeness MEC results of
Refs.~$\cite{mike5,mike6}$ by a factor of between 0.1 and 3.5 (for $\Lambda$
varying over
our reasonable range) over the complete range of $Q^2$ considered in those
calculations.

These results have significant implications for the interpretation of CEBAF
experiment $\cite{cebaf1}$.
Assuming, for example, that the nucleon's strangeness form
factors were identically zero, PV $^4$He asymmetry would still differ from
its \lq\lq zero-strangeness" value [$F_\sst{C}^{(s)}(0)=0$ in
Eq.~(1)] by roughly 15-40\% due to the $\rho$-$\pi$ strangeness MEC.
Thus, for purposes of extracting limits on the nucleon's strangeness electric
form factor, one encounters about a 25\% theoretical uncertainty associated
with non-nucleonic strangeness. By way of comparison, we note that models
which give a large nucleon strangeness electric form factor, $|\GES/\GEn|\sim
1$ generate a 20\% correction to $\alr$ via the one- and two-body mechanims
of Fig. 1a,b. On the other hand, given the projected 40\% experimental error
for the CEBAF measurement, a statistically significant non-zero result
for $F_\sst{C}^{(s)}$
would signal the presence of both a large strange-quark content of the
nucleon as well as a large, non-nucleonic strange-quark component of the
nucleus.

\section{SUMMARY AND CONCLUSIONS}

We have calculated vector-to-pseudoscalar meson vector current form
factors using a combination of vector meson pole and one one-loop
contributions. As expected, the one-loop results are strongly dependent
on the mass parameter in the hadronic form factor needed to regulate
otherwise divergent loop integrals. For values of $\Lambda$ lying
within a \lq\lq reasonable range" whose upper limit is  determined by
self-consistency with
VMD and lower limit by a cloudy bag picture of hadrons, we find that the
one-loop
contributions to the $\rho\pi\gamma$ and $K^\ast K\gamma$ amplitudes may give
a substantial fraction of the measured amplitudes. In the case of
the strange-quark vector current $\rho$-to-$\pi$ transition form
factor, the loop contributions may enhance the total amplitude by more than
a factor of three over an estimate based on VMD. Assuming loops involving
heavier mesonic intermediate states do not cancel the contribution from
the lightest mesons, our results could have serious implications for
the interpretation of the moderate-$|Q^2|$, CEBAF PV electron
scattering experiment with a $^4$He target $\cite{cebaf1}$. Indeed, this
strangeness transition form factor, which contributes to the
nuclear strangeness charge form factor $F_\sst{C}^{(s)}(q)$ via a
meson exchange current $\cite{mike5,mike6}$,
would induce a 15 - 40\% correction to the
zero-strangeness $^4$He PV asymmetry [Eq. (1) with $F_\sst{C}^{(s)} = 0$].
Were the measurement $\cite{cebaf1}$ to extract a statistically significant,
non-zero result for $F_\sst{C}^{(s)}$, one would have evidence that
non-valence quark degrees of freedom -- both nucleonic and
non-nucleonic --- play an important role in the medium energy nuclear response.

\nonumber\section{Acknowledgements}
 J.L.G. was supported by NSF grant HRD-9154080
and by DOE contract  DE-AC05-84ER40150. M.J.M.
was supported by NSF NYI award PHY-9357484 and DOE contracts
DE-AC05-84ER40150 and DE-FG05-94ER0832.
\section{Appendix }
In this appendix we describe the determination of  $R_0$ and  $R_8$.

Determination of $R_0$ and  $R_8$: these two effective couplings can
be determined by
using  the decay width of  $\phi \rightarrow \rho\pi$ and the radiative decays
of
vector mesons supplemented with the hypothesis of VMD.

The partial width $\Gamma(\phi \rightarrow \rho\pi)$ is given by:
\[ \Gamma(\phi \rightarrow \rho\pi)=\frac{\left|
G_{\phi\rho\pi}\right|^2}{12\pi}\;\left| k_f \right|^3 .\]
{}From eq. (13) we obtain:
\[ G_{\phi\rho\pi}=\frac{1}{F_0}\;\left(-R_0 \,\cos\theta
+\frac{4}{\sqrt{3}}\,R_8\,\sin\theta\right),\]
while the experimentally observed partial width gives:
\[\left| G_{\phi\rho\pi}^{{\rm Phen}}\right|=1.08 \;\;{\rm GeV}^{-1}.\]
In practice we will take $\theta$ to correspond to ideal
$\phi-\omega$ mixing.

The radiative transition amplitude $V\rightarrow P\gamma$ has the
general form:
\[A(V\rightarrow
P\gamma)=-i\,g_{VP}^{(\gamma)}\;\epsilon_{\mu\nu\rho\sigma}\,P_V^\mu\,
\varepsilon_V^\nu\,P_\gamma^\rho\,\varepsilon_\gamma^\sigma .\]
The radiative partial width is given by:

\[\Gamma(V \rightarrow P\gamma)= \left|g_{VP}^{(\gamma)}\right|^2\;\left| k_f
\right|^3 .\]
Using VMD we have that:
\[g_{VP}^{(\gamma)}=-\sum_{V^{\prime}=\rho^0, \omega,\phi}\;\frac{G_{V
V^{\prime} P}
C_{V^{\prime}}}{Q^2-M_{V^{\prime}}^2} ,\]
 where $C_V=M_V^2/f_V$, $f_\rho=5.1$, $f_\omega=17$, and  $f_\phi=13$.
{}From eq. (13) we obtain:
\[G_{\rho\rho\pi}=0    \]
\[G_{\rho\omega\pi}=  \frac{1}{F_0} \;\left(
\frac{4}{\sqrt{3}}\,R_8\;\cos\theta+R_0\,\sin\theta  \right)\]
\[G_{\rho\phi\pi}=\frac{1}{F_0} \;\left(
\frac{4}{\sqrt{3}}\,R_8\;\sin\theta-R_0\,\cos\theta\right)\]
\[G_{K^{\ast +}\rho K^+}= \frac{2}{F_0}\;R_8  \]
\[G_{K^{\ast +}\omega K^+}=
\frac{1}{F_0}\;\left(-\frac{2}{\sqrt{3}}\,R_8\,\cos\theta+R_0\,\sin\theta\right)
\]
\[G_{K^{\ast +}\phi K^+}=-\frac{1}{F_0} \;\left(
\frac{2}{\sqrt{3}}\,R_8\,\sin\theta+R_0\,\cos\theta \right)\]
\[G_{K^{\ast 0}\rho K^0}= -G_{K^{\ast +}\rho K^+}\]
\[G_{K^{\ast 0}\omega K^0}=G_{ K^{\ast +}\omega K^+ } \]
\[G_{K^{\ast 0}\phi K^0}= G_{K^{\ast +}\phi K^+}  \]
Using the measured rates, our best fit leads to:
\[R_0\sim 1.0~~~~,~~~~~~~R_8\sim0.27.\]

The VMD result for the transition matrix elements of the strangeness current
have the same structure, the only difference is that now
$C_V$ has to be replaced by $S_V$. In the following we assume that
$S_V$ is only non-vanishing for $V=\phi$. This holds whenever $C_0$
corresponds to exact OZI suppression as mentioned in section III.
{}From eq. (16) we obtain:
\[S_{\phi}=-  3 \;\frac{M_\phi^2}{f_\phi}.\]
This leads to the VMD result used in the text:
\[\frac{1}{M_\rho}\,g_{\rho \pi}^{(s)}(Q^2=0)\Bigr\vert_{\phi-{\rm dom}}=
 \frac{1}{M_\phi^2}\, G_{\rho \phi\pi}^{{\rm Phen}} S_{\phi} .\]

\def\prd#1#2#3{{\it Phys. ~Rev. ~}{\bf D#1} (19#2) #3}
\def\prc#1#2#3{{\it Phys. ~Rev. ~}{\bf C#1} (19#2) #3}
\def\plb#1#2#3{{\it Phys. ~Lett. ~}{\bf B#1} (19#2) #3}
\def\npb#1#2#3{{\it Nucl. ~Phys. ~}{\bf B#1} (19#2) #3}
\def\npa#1#2#3{{\it Nucl. ~Phys. ~}{\bf A#1} (19#2) #3}
\def\prl#1#2#3{{\it Phys. ~Rev. ~Lett. ~}{\bf #1} (19#2) #3}
\def\prep#1#2#3{{\it Phys. ~Rep. ~}{\bf #1} (19#2) #3}
\def\Zphysc#1#2#3{{\it Z. ~Phys. ~}{\bf C#1} (19#2) #3}
\bibliographystyle{unsrt}

\begin{thebibliography}{99}
\bibitem{correct} The value quoted in Refs. $\cite{mike5,mike6}$ should
be corrected by multiplying by $\mro/m_\phi\approx 0.75$ to account
properly for the normalization of the transition and $\phi$-decay
matrix elements.
\end{thebibliography}

\newpage

\begin{center}
{\bf FIGURE CAPTIONS}
\end{center}

 {{\bf Figure 1:}  {Contributions to the nuclear
strangeness charge form factor, $\fcnucs (q)$. One-body (a) and
\lq\lq pair current" (b) contributions depend on the nucleon's
strange-quark vector current form factors. \lq\lq Transition current" (c)
contributions arise from strange-quark vector current matrix elements
between meson states $\ket{M}$ and $\ket{M'}$. Here, the cross indicates
the insertion of the strangeness charge operator. }}

\vspace*{0.5in}

 {{\bf Figure 2:}  {$\phi$-meson dominance picture of the
strangeness current
 transition matrix element. $V$ refers to a vector meson, $P$ to a
pseudoscalar,
and the cross represents the insertion of the current.}}

\vspace*{0.5in}

 {{\bf Figure 3:}  {  One-loop diagrams which contribute to the
strangeness current transition matrix element.
The cross represents the strangeness vector current.}}

\vspace*{0.5in}

 {{\bf Figure 4:}  { Ratio $\rgamropi$ for
the matrix element of the electromagnetic current
between $\rho$ and $\pi$ mesons. The solid curve correspond to case
1 ($\zeta=0.16$) and the dashed one to case 2 ($\zeta=-0.08$).
In each case we have used  $k_1^2=M_{\rho}^2$, $Q^2=0$, and $k_2^2=M_\pi^2$.}}

\vspace*{0.5in}

 {{\bf Figure 5:}  {  Ratio $\rgamkk$ for the matrix
element of the electromagnetic current
between $K^\ast$ and $K$ mesons. The solid lines correspond to  charged
kaons and the dashed ones to neutral kaons.
In each case, the upper curve corresponds to case 2, and the lower one to case
1. The kinematic invariants used are:
  $k_1^2=M_{K^\ast}^2$, $Q^2=0$, and $k_2^2=M_K^2$.}}

\vspace*{0.5in}

 {{\bf Figure 6:}  {  Ratio $\rsropi$ for the matrix
element of the strangeness current between $\rho$ and $\pi$ mesons.
The solid curves corresponds to case 1 and the dashed ones to case 2.
In each case the lower curve  corresponds to
 $k_1^2=M_{\rho}^2$, $Q^2=0$ and $k_2^2=M_\pi^2$, and the upper one to
$k_1^2=0$, $Q^2=0$ and $k_2^2=0$. The dash-dotted  curve corresponds to
case 1 with  $k_1^2=M_{\rho}^2$, $Q^2=0.6$ $ {\rm GeV^2}$ and $k_2^2=M_\pi^2$,
and the dotted curve corresponds to case 1 with $k_1^2=M_{\rho}^2$, $Q^2=0$ and
$k_2^2=M_\pi^2$, and the form factor $F(k^2)=\Lambda^2/(\Lambda^2-k^2)$. }}

\end{document}